\documentclass[final,5p,times,twocolumn]{elsarticle}

\usepackage{picins}

\usepackage{lineno,multirow,diagbox,multicol,balance,epstopdf,amsmath, subfigure}
\usepackage{amssymb,amsfonts}
\usepackage{algorithmic}
\usepackage{float}
\usepackage{stfloats}
\usepackage{subfigure}
\usepackage{amssymb,amsmath,bm}
\usepackage{soul}
\usepackage{tabularx}
\usepackage{textcomp}
\usepackage{multirow}
\usepackage{caption}
\usepackage{booktabs}
\usepackage{makecell}
\usepackage{color}
\usepackage[colorlinks,linkcolor=blue,anchorcolor=blue,citecolor=blue,CJKbookmarks=True]{hyperref}
\usepackage{url}

\journal{Neural Networks}

\begin{document}

\begin{frontmatter}

\title{Noise-robust voice conversion with domain adversarial training}

\author[1,3]{Hongqiang Du}
\ead{hqdu@nwpu-aslp.org}

\author[1]{Lei Xie\corref{mycorrespondingauthor}}
\cortext[mycorrespondingauthor]{Lei Xie is the corresponding author.}

\author[2,3]{Haizhou Li}



\address[1]{Audio, Speech and Language Processing Group (ASLP@NPU), School of Computer Science, \\Northwestern Polytechnical University, Xi'an, China}

\address[2]{The Chinese University of Hong Kong (Shenzhen), China}

\address[3]{Department of Electrical and Computer Engineering, National University of Singapore, Singapore}


\begin{abstract}

Voice conversion has made great progress in the past few years under the studio-quality test scenario in terms of speech quality and speaker similarity. However, in real applications, test speech from source speaker or target speaker can be corrupted by various environment noises, which seriously degrade the speech quality and speaker similarity. In this paper, we propose a novel encoder-decoder based noise-robust voice conversion framework, which consists of a speaker encoder, a content encoder, a decoder, and two domain adversarial neural networks. Specifically, we integrate disentangling speaker and content representation technique with domain adversarial training technique.  Domain adversarial training makes speaker representations and content representations extracted by speaker encoder and content encoder from clean speech and noisy speech in the same space, respectively. In this way, the learned speaker and content representations are noise-invariant. Therefore, the two noise-invariant representations can be taken as input by the decoder to predict the clean converted spectrum. The experimental results demonstrate that our proposed method can synthesize clean converted speech under noisy test scenarios, where the source speech and target speech can be corrupted by seen or unseen noise types during the training process. Additionally, both speech quality and speaker similarity are improved.

\end{abstract}
\end{frontmatter}

\section{Introduction}
Voice conversion (VC) is a technique to transform the speech signal of a source speaker to sound like that of a target speaker without changing the linguistic content~\cite{mohammadi2017overview}. This technique has many applications, including voice morphing, emotion conversion, speech enhancement~\cite{mouchtaris2004spectral}, movie dubbing as well as other entertainment applications.

Voice conversion has taken some major strides in terms of speech quality and speaker similarity. Various approaches have been proposed, such as Gaussian
mixture model (GMM)~\cite{benisty2011voice,stylianou1998continuous,toda2007voice}, frequency warping approaches~\cite{erro2009voice,godoy2011voice,tian2015sparse}, exemplar based methods~\cite{takashima2012exemplar,wu2014exemplar,tian2017exemplar}, and neural network based methods~\cite{sun2015voice,hsu2016voice,hsu2017voice,kaneko2017parallel,kaneko2018cyclegan,kameoka2018stargan,tanaka2019atts2s,zhang2019sequence,du2021optimizing,wang2021enriching}. 
Recently, disentangling speaker and linguistic content representations based on deep learning for voice conversion~\cite{chou2019one,qian2019autovc,du2021improving,wang2021one} has received much attention. Comparing with conventional methods, source speakers and target speakers during evaluation are not required to be seen in the training process. Disentangling approaches achieve good performance in terms of speech quality and speaker similarity when evaluated under a clean scenario, where both source speech and target speech are clean.

Despite recent progress, the test speech from source and target speaker can be corrupted by various environment noises in real applications. Recent studies~\cite{takashima2012exemplar,huang2021far} show that many popular voice conversion frameworks, including disentangling methods:  AdaIN-VC~\cite{chou2019one}, and AUTOVC~\cite{qian2019autovc}, DGAN-VC~\cite{chou2018multi}, suffer serious speech quality and speaker similarity degradation in noisy conditions. Corpora from studio settings are greatly different from real-world testing conditions. In noisy conditions, linguistic content, speaker identity, and noise are intermingled together in speech signals. How these factors are intermingled to compose the speech signal and impact each other are far from clear~\cite{li2018deep}. Due to the complex nature of the noise corruption process, the linguistic content representations and speaker representations from clean speech and corresponding noise corrupted speech have different distributions,  respectively~\cite{sun2017unsupervised,wang2018unsupervised}.

There have been a few techniques to address such problem in voice conversion. Non-negative matrix factorization (NMF)~\cite{takashima2012exemplar,aihara2014noise,aihara2015small} assumes that the speech can be expressed with exemplars and corresponding weights. NMF builds a dictionary consisting of corresponding exemplars from source speech and target speech. The converted spectrum is reconstructed with target exemplars and the picked weights related to the source exemplars. However, this method only considers background noise in source speech and target speech should be clean during evaluation. Recently, Hsu et al.~\cite{hsu2017unsupervised} proposed to learn disentangled latent representations. Hsu et al. explored a hierarchical latent space which encodes different attributes into latent segment variables and latent sequence variables. At run-time, by replacing the latent sequence variable from noise speech with that of a clean utterance, this framework is able to synthesize denoised converted speech. However, this method requires that the target speech is clean. Therefore, it remains a challenge to generate clean converted speech under complex noisy conditions in real applications, where both source and target speech are corrupted by various noises.

To build a noise-robust voice conversion system that is robust to complex noisy conditions, a straightforward idea is to use speech enhancement as a pre-processing module~\cite{valentini2016investigating} to get denoised speech for downstream tasks. But the inevitable distortion of denoised speech can lead to clear quality deterioration to the synthesized speech. This conclusion has been further confirmed in~\cite{yang2020adversarial}. Additionally, the method strongly relies on prior knowledge of noise~\cite{sekkate2019investigation}, which limits their applications. Hsu et al.~\cite{hsu2019disentangling} explored adversarial training for disentangling speaker attribute from noise attribute. This technique can independently control the speaker identity and background noise in the generated speech.  
Learning noise-invariant features is another successful attempt. Domain adversarial training (DAT) is a popular method to extract domain-invariant representations. Domain adversarial neural network consists of a feature extractor, a gradient reversal layer (GRL)~\cite{ganin2016domain}, a task classifier, and a domain classifier. The feature extractor extracts a representation that is discriminative to the task classifier, while is indiscriminate to the domain classifier with the help of GRL, which is referred as domain-invariant representation. Recently, due to easy implementation and great performance, DAT has been applied in speech recognition~\cite{sun2017unsupervised,shinohara2016adversarial}, speaker verification~\cite{wang2018unsupervised,tu2019variational}, speech enhancement~\cite{liao2018noise} and wake-up word detection~\cite{lim2020cross}.


Building on the success of the prior studies on  
disentangling content and speaker representations and domain adversarial training, in this paper, we propose a novel encoder-decoder based noise-robust voice conversion method. AdaIN-VC~\cite{chou2019one} is a successful end-to-end disentangling approach, which consists of a speaker encoder, a content encoder, a decoder, and they are jointly optimized. Compared with GAN~\cite{kaneko2018cyclegan,kameoka2018stargan} based and sequence to sequence~\cite{tanaka2019atts2s,zhang2019sequence} based methods, AdaIN-VC can perform non-parallel~\cite{hsu2016voice} any-to-any voice conversion. Moreover, Huang et al.~\cite{huang2021far} confirmed that AdaIN-VC is more robust than DGAN-VC and AUTOVC in noisy conditions. Therefore, we use AdaIN-VC as a case study. Based on successful framework design of AdaIN-VC, we extend the content encoder and speaker encoder with a gradient reversal layer (GRL) and a domain classifier, respectively. Note that we do not need the phone or speaker related task classifier to extract content and speaker representations. The gradient reversal layer helps to reduce the performance of the domain classifier. At a result, the gap between representations extracted from noisy domain and clean domain also reduced. Domain adversarial training makes the learned content and speaker representations from noisy speech and clean speech noise-invariant, respectively. In this way, the decoder can synthesize clean converted speech with noise-invariant content representation from source speech and speaker representation from target speech when either source or target speech is corrupted by noises at run-time.

The main contributions are as follows: first, we learn noise-invariant content representation and speaker representation for voice conversion in an unsupervised manner. Second, since the available works concentrate on only source speech is corrupted by noise at run-time, this paper represents a new effort to approach real-life conditions where the utterances from source and target speaker may be corrupted by various noises. Third, we find that our proposed method is robust for different noise types. Even for unseen noise types, our system can synthesize clean converted speech.

The rest of the paper is organized as follows. In Section~\ref{sec:related work}, we briefly discuss the related work, including disentangling content and speaker representations for voice conversion, and domain adversarial training. In Section~\ref{sec:proposed}, we introduce our proposed noise-robust voice conversion method. In Section~\ref{sec:experiment setups}, we introduce the experiment setups. In Section~\ref{sec:experiment results}, we report and analyze the experimental results. We conclude this paper in Section~\ref{sec:conclusion}.

\section{Related Work}
\label{sec:related work}

In this section, we will give an overview of disentangling content and speaker representations for voice conversion, and domain adversarial training, to set the stage for our study.

\subsection{Disentangling content and speaker representations for voice conversion}
\label{sec:disentangling}

\begin{figure}[!ht]
	\centering
	\centerline{\includegraphics[width=1.0\linewidth]{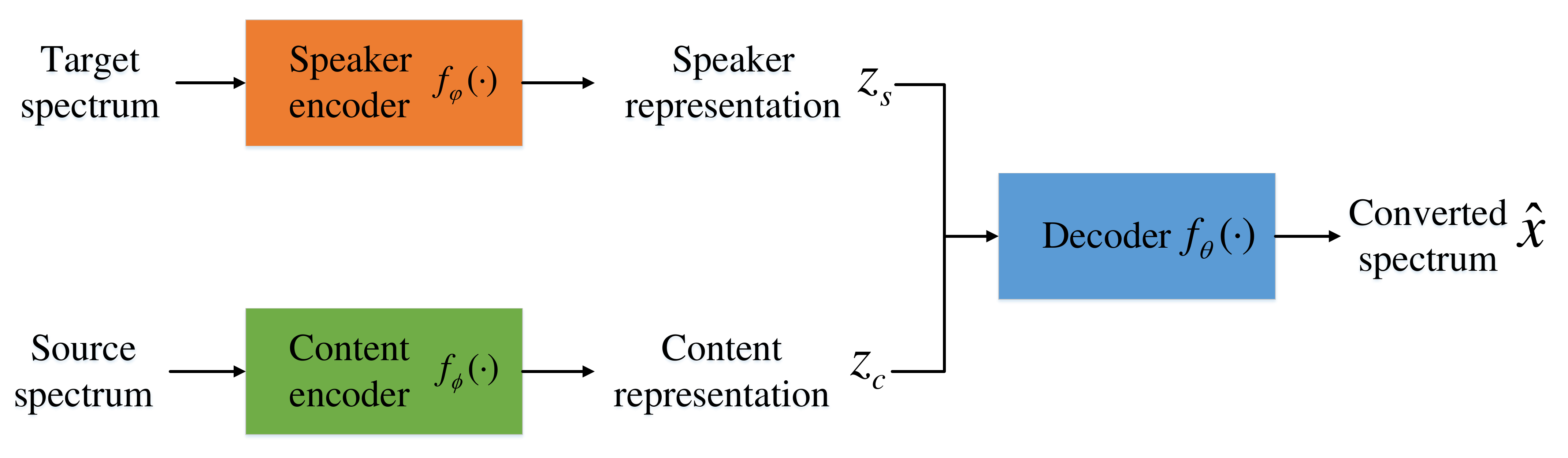}}
	\caption{Run-time process of voice conversion framework based on disentangling content and speaker representations. The speaker encoder extracts speaker representation from target speech. The content encoder extracts content representation from source speech. The decoder takes content and speaker representations as input to reconstruct converted spectrum.}
	\label{fig:general}
\end{figure}

The speech signal can be factorized into a speaker representation and a content representation, meanwhile speech signals can also be recovered by the two explanatory factors of variation~\cite{chou2019one}. Voice conversion asks to maintain content information while changing the speaker information in one utterance. Figure~\ref{fig:general} shows the run-time process of voice conversion framework based on disentangling content and speaker representations. The speaker encoder ${f_\varphi }( \cdot )$ takes spectrum from target speech as input and the output is speaker representation ${z_s}$ in utterance level. The content encoder ${f_\phi }( \cdot )$ takes spectrum from source speech as input and the output is linguistic content representation ${z_c}$ in frame level. The decoder utilizes the concatenated latent representation $({z_s},{z_c})$ to reconstruct converted spectrum. In this way, we can control the speaker identity in the generated speech independently. The conversion process is formulated as follows, where $\widehat x$ is the converted spectrum.

\begin{equation}\label{eq:rec}
	\widehat x = {f_\theta }({z_s},{z_c})
\end{equation}

AdaIN-VC~\cite{chou2019one} adopts the same framework structure as Figure~\ref{fig:general} for voice conversion. The speaker and content representations can be extracted successfully by designing the neural network of speaker encoder and content encoder.

Speaker identity is time-independent and merely changes in one utterance, hence it is static~\cite{chou2019one}. To disentangle speaker representation from spectrum, the speaker encoder first uses 1D convolution layers to get frame-level speaker features. Then it adopts average pooling layers to aggregate frame-level speaker information to form utterance level speaker representation~\cite{okabe2018attentive,tang2019deep}.

Linguistic content changes dramatically among frames in one utterance, hence it is dynamic~\cite{chou2019one}. It is important for the content encoder not to memorize the input spectrum but to encode it in a semantic way~\cite{mor2018universal}. The content encoder compresses the spectrum to form bottleneck representation, which means that a part of the information, e.g. speaker information, is lost during the compressing process and the remaining information should provide the decoder with sufficient information that is necessary for perfect reconstruction~\cite{qian2019autovc}. To further remove speaker related information, the content encoder also utilizes instance normalization technique~\cite{ulyanov2016instance}. Finally, a multivariate content representation is sampled from bottleneck representation, which corresponds to semantically meaningful factors of variation of the observations (e.g., linguistic content)~\cite{hsu2016voice,locatello2019challenging}.

To sum up, disentangling content and speaker representations for voice conversion requires that the speaker encoder and content encoder are designed to discard unnecessary information while preserving relevant information for the task of interest. The decoder attempts to reconstruct the spectrum from the compressed representations extracted by the content encoder and speaker encoder.

\subsection{Domain adversarial training}

Domain adversarial training (DAT) is designed to learn domain-invariant representation by reducing the bias presented in data from different domains~\cite{ganin2016domain}. The framework of domain adversarial neural network is shown in Figure~\ref{fig:dat}, which consists of a feature extractor, a gradient reversal layer (GRL), a task classifier, and a domain classifier. The key component is the gradient reversal layer. Learning domain-invariant representation is achieved by inserting a gradient reversal layer between the feature extractor and the two classifiers: task classifier and domain classifier. The gradient reversal layer (GRL) can be formulated as Eq~\ref{eq:grl}:

\begin{figure}[!ht]
	\centering
	\centerline{\includegraphics[width=1.0\linewidth]{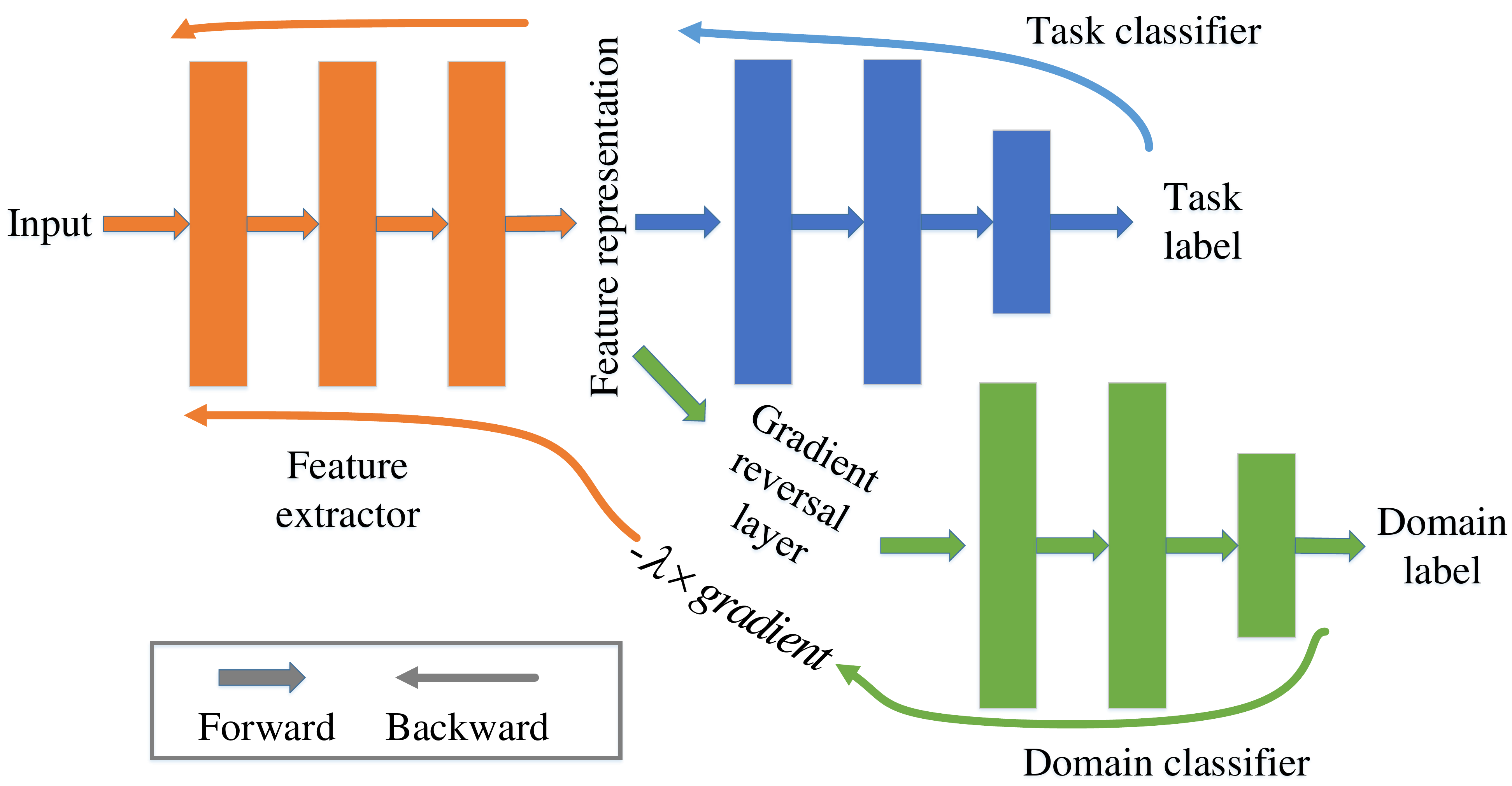}}
	\caption{The framework of domain adversarial neural network. The network consists of four components: a feature extractor, a gradient reversal layer, a task classifier, and a domain classifier.}
	\label{fig:dat}
\end{figure}

\begin{equation}\label{eq:grl}
	{\rm{GRL}}(x) = \left\{ {\begin{array}{*{20}{c}}
			{x,}&{forward}\\
			{ - \lambda  \times gradient,}&{backward}
	\end{array}} \right.
\end{equation}
where $x$ is input, $\lambda$ is a scaling factor of the gradient. In the forward-pass, the GRL acts as an identity layer that leaves the input unchanged. In backward-pass, the gradient is multiplied by a negative scalar $\lambda$ and propagated back to the shared feature extractor layers. The whole network is optimized to minimize the error of task classifier while maximizing the error of the domain classifier with the help of GRL~\cite{sun2017unsupervised,wang2018unsupervised,ganin2016domain}. The low performance of domain classifier indicates that the gap between learned representations from different domains is reduced.  As a result, the extracted representations from different domains are projected into the same feature subspace, and they have very similar distributions~\cite{wang2018unsupervised}, which are indistinguishable to the domain classifier but are discriminative to the task classifier. The downstream tasks will not take domain information into consideration, and thus will be robust to domain variation.

\section{Noise-robust voice conversion with DAT}
\label{sec:proposed}
In this section, we will introduce our proposed noise-robust voice conversion method, which integrates disentangling content and speaker representations technique, and domain adversarial training technique.

\subsection{Disentangling noise-invariant content and speaker representations with DAT}
Domain adversarial training has been successfully applied to speech recognition~\cite{shinohara2016adversarial,sun2017unsupervised} and speaker verification~\cite{wang2018unsupervised}. For speech recognition task, the feature extractor extracts domain-invariant content representation by minimizing the loss of the phoneme classifier while maximizing the loss of the domain classifier. For speaker verification task, the feature extractor extracts domain-invariant speaker representation by minimizing the loss of speaker classifier while maximizing the loss of the domain classifier. But for our work, we extract noise-invariant speaker and content representations in an unsupervised  manner. We do not add explicitly phoneme classifier and speaker classifier. 

Previous work~\cite{hsu2019disentangling} adopts noise augmentation and domain adversarial training in text to speech (TTS). Our work substantially differs from the above work in following aspects: (1) VC is a related but different task from TTS. For TTS, the input is text and the output is spectrum. For VC, both input and output are spectrum. (2) The work~\cite{hsu2019disentangling} adopts noise augmentation and domain adversarial training mainly for extracting noise-invariant speaker embedding, while our work focuses on extracting noise-invariant speaker embedding and content embedding. (3) To extract noise-invariant representations, we only use domain classifier, while previous work uses both task and domain classifiers. Furthermore, domain adversarial training is separately used for each element of content representation in our work. (4) In the previous work, noise and speaker are correlated, and speech of one speaker is augmented with random SNRs to make SNRs less discriminative about speakers, while in our work, we consider more practical situations -- the speech of one speaker is corrupted with various noise types at a random SNR to make learned representation robust against different noise types and SNRs. Next, we will introduce our proposed method.

\begin{figure*}[!ht]
	\centering
	\centerline{\includegraphics[width=0.9\linewidth]{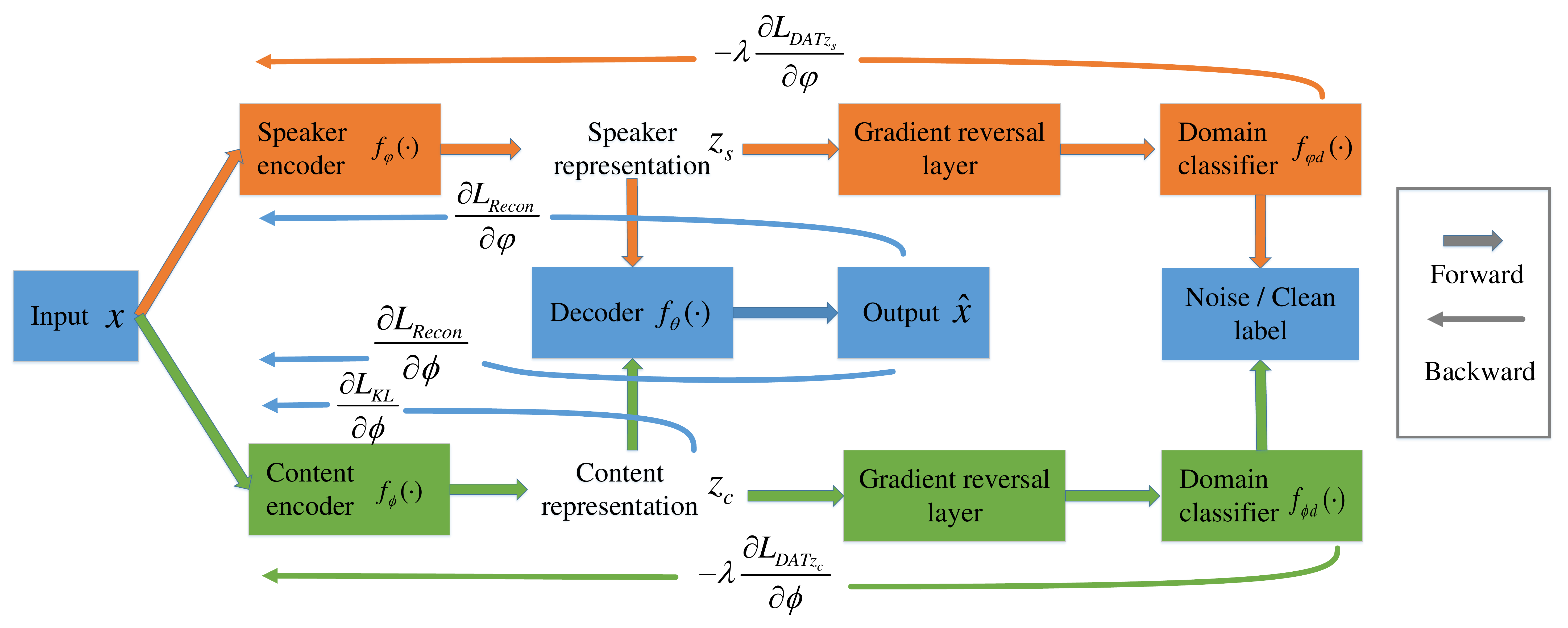}}
	\caption{The diagram of our proposed noise-robust voice conversion method. This framework consists of a speaker encoder ${f_\varphi }( \cdot )$, a content encoder ${f_\phi }( \cdot )$, two gradient reversal layers and two domain classifiers ${f_{\varphi d}}( \cdot )$,  ${f_{\phi d}}( \cdot )$ for speaker encoder and content encoder,  and a decoder ${f_\theta }( \cdot )$.}
	\label{fig:framework}
\end{figure*}

We denote the training speech dataset as ${\rm{\{ }}{x_i}{\rm{,}}{d_i}{\rm{\} }}_{i = 1}^N$. $x_i$ is the $i$-th data sample, ${d_i}$ is the corresponding domain label. ${d_i} = 0$ indicates that ${x_i}$ comes  from  the  clean domain,  ${d_i} = 1$ represents that ${x_i}$ is from noisy domain. Note that we do not use explicit noise types, e.g. street, cafe, as noisy domain labels. We use ${d_i}$ as domain-label to make the extracted representations noise-robust to different noise types and SNRs. The data from clean domain and noisy domain are paired. Our goal is to utilize these clean and noisy data to learn noise-invariant linguistic content representation and speaker representation for noise-robust voice conversion.

The framework of our proposed noise-robust voice conversion method is shown in Figure~\ref{fig:framework}. It consists of following components: a speaker encoder ${f_\varphi }( \cdot )$, a content encoder ${f_\phi }( \cdot )$, two gradient reversal layers and two domain classifiers ${f_{\varphi d}}( \cdot )$,  ${f_{\phi d}}( \cdot )$ for speaker encoder and content encoder, and a decoder ${f_\theta }( \cdot )$. Here, $\phi$, $\varphi$, $\phi d$, $\varphi d$ and $\theta$ are the parameters of the corresponding network. The whole network shown in Figure~\ref{fig:framework} can be formulated as follows:

\begin{equation}\label{eq:content}
	{z_c} = {f_\phi }(x)
\end{equation}

\begin{equation}\label{eq:speaker}
	{z_s} = {f_\varphi }(x)
\end{equation}

\begin{equation}\label{eq:content_domain}
	{d_{oc} = {f_{\phi d}}({z_c})}
\end{equation}

\begin{equation}\label{eq:speaker_domain}
	{d_{os} = {f_{\varphi d}}({z_s})}
\end{equation}

\begin{equation}\label{eq:domain}
	\hat x = {f_\theta }({z_s},{z_c})
\end{equation}
where the input $x$ for speaker encoder and content encoder may be a noisy or clean utterance randomly selected from ${\rm{\{ }}{x_i}{\rm{,}}{d_i}{\rm{\} }}_{i = 1}^N$. ${z_s}$ and ${z_c}$ are the extracted noise-invariant speaker and content representation,  respectively. $d_{oc}$ and $d_{os}$ are outputs of the two domain classifiers. The output $\hat x$ of decoder is the clean converted spectrum. Our proposed model reconstructs clean speech from noisy speech, which makes the model like a denoising autoencoder~\cite{lu2013speech,shivakumar2016perception}.

The proposed framework for noise-robust voice conversion consists of two steps: conversion model training and run-time conversion. The forward pass of training process is shown in Figure~\ref{fig:framework}. On the one hand, the speaker encoder and content encoder extract speaker representation and content representation from clean or noisy speech. The learned representations from clean and noisy speech have different distributions. To make the representations ${z_s}$ and ${z_c}$ noise-invariant, two gradient reversal layers take ${z_s}$ and ${z_c}$ as input, respectively. Then the outputs of two gradient reversal layers are fed into two domain classifiers, respectively. As there are two domains, the domain classifiers are binary classifiers.  On the other hand, the decoder takes the speaker and content representations to reconstruct the clean spectrum.  The backward pass of training process is shown in Figure~\ref{fig:framework}. ${L_{Recon}}$, ${L_{KL}}$, ${L_{DAT{z_s}}}$, and ${L_{DAT{z_c}}}$ are four loss functions. The details of the loss functions will be introduced in Section~\ref{subsec:loss_function}. The parameters $\varphi$ of speaker encoder are updated by ${L_{Recon}}$ and ${L_{DAT{z_s}}}$. The parameters $\phi$ of content encoder are updated by ${L_{Recon}}$, ${L_{KL}}$ and  ${L_{DAT{z_c}}}$. The gradients in two gradient reversal layers are multiplied by a negative scalar $\lambda$ and propagated back to the content encoder and speaker encoder, respectively. In this way, the learned content and speaker representations make the error of two domain classifiers increase. The gap of distributions between representations from noisy speech and clean speech is reduced, which helps to project the speaker and content representations into a noise-invariant subspace,  respectively.  In this subspace, the learned speaker and content representations satisfy the following condition:

\begin{equation}\label{eq:domain1}
	f({x_{i,{d_i} = 0}}) = f({x_{i,{d_i} = 1}})
\end{equation}

In this way, the domain information is not included in the learned representations. Therefore, the decoder can take noise-invariant speaker and content representations as input to reconstruct the clean speech. 

During run-time process, only the speaker encoder, content encoder and decoder work. The inputs of content encoder and speaker encoder are from source speaker and target speaker, respectively. Then the learned noise-invariant speaker representation from target speaker and content representation from source speaker are fed into decoder to get clean converted output. As the speech from source speaker and target speaker may be clean or noisy, our proposed noise-robust voice conversion method can achieve voice conversion under four scenarios: (1) both source speech and target speech are clean (SC-TC); (2) source speech is clean while target speech is noisy (SC-TN); (3) source speech is noisy while target speech is clean (SN-TC); (4) both source and target speech are noisy (SN-TN).

\subsection{Loss functions}
\label{subsec:loss_function}
Our proposed noise-robust voice conversion framework is jointly optimized. The overall loss function is a linear combination of reconstruction loss, Kullback-Leibler (KL) loss, and two domain classification losses.  During the training process, we seek to minimize the reconstruction loss and KL loss while maximizing the two domain classification losses. The formula is expressed as follows:

\begin{equation}\label{eq:loss}
	\begin{aligned}
		L(\phi ,\varphi ,\phi d,\varphi d,\theta ) &= \alpha {L_{Recon}}(\phi ,\varphi ,\theta ) + \beta {L_{KL}}(\phi ) \\& + \tau {L_{DAT{z_c}}}(\phi ,\phi d) \\& +  \gamma {L_{DAT{z_s}}}(\varphi ,\varphi d)
	\end{aligned}
\end{equation}
where $\alpha$, $\beta$, $\gamma$, $\tau$  are the hyper-parameters to balance different losses. The first term ${L_{Recon}}$ is the reconstruction loss, which is a  mean absolute error between reconstructed spectrum and clean spectrum. The second term ${L_{KL}}$ is the Kullback-Leibler (KL) divergence loss between the content representation’s posterior $q(z_c|x)$ and prior $p(z_c)$. The prior is assumed to be a centered isotropic multivariate Gaussian $p({z_c}) = {\rm N}(z_c;0,I)$, where $I$ is the identity matrix. Minimizing the KL loss encourages content encoder to learn linguistic content representation~\cite{hsu2016voice}. Minimizing the reconstruction loss encourages the decoder to learn to reconstruct the input with content and speaker representations.

The third term $L_{DAT{z_c}}$ and the fourth term ${L_{DAT{z_s}}}$ are the two domain classifier losses. As we want to make content and speaker representations noise-invariant, the learned representations ${z_c}$ and ${z_s}$ should make the well-trained two domain classifiers  ${f_{\phi d}}( \cdot )$ and  ${f_{\varphi d}}( \cdot )$ fail to
distinguish which domain the representation comes from. To achieve this,  the two domain classification loss ${L_{DAT{z_c}}}(\phi ,\phi d)$  and ${L_{DAT{z_s}}}(\varphi ,\varphi d)$ are maximized.

Overall, by using the four loss functions, the learned content and speaker representations can keep enough relative information for perfect reconstruction. Meanwhile, they are robust to domain variations.

\section{Experimental setups}
\label{sec:experiment setups}
To validate our proposed method, we develop three proposed systems with different configurations: domain adversarial training is used only for speaker encoder, content encoder, and both speaker encoder and content encoder, respectively. Then we select our best system and compare it with other baseline systems under four test scenarios: both source speech and target speech are clean
(SC-TC); (2) source speech is clean and target speech is
noisy (SC-TN); (3) source speech is noisy while target speech
is clean (SN-TC); (4) both source and target speech are noisy
(SN-TN).

\subsection{Database and feature extraction}

CSTR-VCTK database~\cite{veaux2017cstr} is a clean multi-speaker corpus, which contains 44 hours of speech samples from 109 speakers. A noise corpus from CHiME4 challenge~\cite{vincent2017analysis} contains about 8.5 hours of background noises recorded in four different locations (bus, cafe, pedestrian area, and street), was used to simulate noisy speech. Three-fourths of noise corpus were used for training and the rest for testing. Each utterance from VCTK corpus was corrupted with four noise types at a random signal-to-noise ratio (SNR) ranging from 5dB to 20dB. The clean and augmented speech together were used as the training dataset to train the voice conversion model.

Voice conversion experiments were carried out on CMU-ARCTIC~\cite{kominek2004cmu} database. We selected two female and two male speakers, and the following conversion pairs were conducted: female to male, female to female, male to male, male to female.  Usually, a small number of utterances~\cite{tian2017exemplar} for each speaker are used for test. In this paper, 30 utterances from each conversion pair were used for evaluation. 120 converted utterances were synthesized for the four speaker pairs in total. All audio files were downsampled to 16 kHz. To test performance of our proposed method, we conducted experiments under seen and unseen noise conditions. For the seen noise conditions, we used the remaining noise clips from CHiME-4 to simulate noisy test speech at 5 dB, 10 dB, 15dB, and 20 dB SNR. For the unseen noise condition, we selected \emph{babble and hfchannel} noises from NOISEX-92~\cite{varga1993assNOISEessment} corpus and added them to the test speech at 5 dB, 10 dB, 15dB, and 20 dB SNR, respectively. A noisy speech corpus CSTR-NOISE~\cite{botinhao2016investigating} was used to test different systems under real noisy conditions. The noise conditions include cafe, restaurant, car, kitchen and meeting room. We selected two female and two male speakers, 30 utterances of each speaker were utilized for evaluation.

Mel spectrogram is a compact representation of the audio signal. Librosa~\cite{mcfee2015librosa} was employed to extract 256 dimensional mel spectrogram from clean and noisy speech with 50ms frame length and 12.5ms frame shift. Neural vocoder Parallel WaveGAN~\cite{yamamoto2020parallel} was used to synthesize the converted speech.

\subsection{System architectures}
\label{subsec:systems}

The details of baselines and the proposed noise-robust voice conversion systems are introduced as follows. All the following methods are conducted for any-to-any voice conversion. As NMF is a traditional one-to-one voice conversion method, we did not use it as a baseline.

\begin{enumerate}[1.]
	\item Baseline systems
	\begin{enumerate}[\textbullet]
		\item VAE-C-C: AdaIN-VC~\cite{chou2019one} system takes 256 dimensional mel spectrogram from clean speech as input and the output is 256 dimensional clean mel spectrogram.
		\item VAE-CN-C: VAE-CN-C has the same setting as VAE-C-C except that this takes 256 dimensional mel spectrogram from clean or noisy speech as input and the output is 256 dimensional clean mel spectrogram.
		\item VAE-CD-C: VAE-CD-C has the same setting as VAE-CN-C except that this system takes mel spectrogram from clean or denoised speech as input. For this denoising baseline system, we use the state-of-the-art speech enhancement model named DCCRN~\cite{hudccrn} to get denoised speech. 
		\item FHVAE-CN-CN: FHVAE-CN-CN~\cite{hsu2017unsupervised} takes 256 dimensional  mel spectrogram from clean or noisy speech as input and the output is 256 dimensional clean or noisy mel spectrogram. Note that this system only works in one kind of noisy scenario where the source speech is noisy and target speech is clean.
	\end{enumerate}

	\item Proposed systems
	\begin{enumerate}[\textbullet]
		\item VAEDC-CN-C: This is one of our proposed noise-robust voice conversion systems. Domain adversarial training is only used for the content encoder to extract noise-invariant content representation. This system takes 256 dimensional mel spectrogram from clean or noisy speech as input and the output is 256 dimensional clean mel spectrogram.

		\item VAEDS-CN-C: VAEDS-CN-C has the same setting as VAEDC-CN-C except that domain adversarial training is only used for the speaker encoder to extract noise-invariant speaker representation. 
		
		\item VAED-CN-C: VAED-CN-C has the same setting as VAEDC-CN-C except that domain adversarial training is used for both content encoder and speaker encoder to extract noise-invariant content and speaker representations.
	\end{enumerate}
	
\end{enumerate}

AdaIN-VC system adopts an encoder-decoder based framework. We make some adjustments to the speaker encoder and decoder to improve performance. The speaker encoder consists of a ConvBank block~\cite{chou2019one}, a residual network (ResNet) and a dense block~\cite{chou2019one}. The content encoder follows original configurations. The speaker encoder and content encoder take 256 dimensional mel spectrogram as input and the output is 128 dimensional speaker and content representations, respectively. To further improve the speech quality, auto-regressive technique is used in the decoder. The domain adversarial neural network consists of a gradient reversal layer, a dense layer and a softmax layer. The whole network is optimized with Adam optimizer~\cite{2014Adam}. The scalar $\lambda$ is set to 0.1.  The hyper-parameters $\alpha$, $\beta$, $\gamma$, $\tau$ are set to 10, 0.5, 0.1, 0.1, respectively.

Speech enhancement model DCCRN~\cite{hudccrn} ranked first for the real-time-track and second for the non-real-time track in terms of Mean Opinion Score (MOS). We used an internal DCCRN model. The noisy speech was simulated with dynamic mixing during model training. The total data seen by DCCRN was over 2000 hours. For the noise-robust voice conversion model FHVAE, we used the original neural network configurations~\cite{hsu2017unsupervised}.

\subsection{Evaluation metrics}
Both objective and subjective evaluations were conducted to evaluate the systems.

\subsubsection{Objective evaluation}

Mel-cepstral distortion (MCD)~\cite{toda2007voice} was employed to measure the spectral distortion. Euclidean distance of the 40 dimensional mel cepstral coefficients (MCC) between the converted speech and the
target speech is calculated with MCD formula. Given a speech frame, the MCD is defined as follows: 
\begin{equation}\label{eq:mcd}
	\vspace{1mm}
	\text{MCD[dB]} = \frac{{10}}{{\ln 10}}\sqrt {2\sum\limits_{n=1}^N {{{\left( {X_n^{conv} - X_n^{targ}} \right)}^2}} },
\end{equation}
where $X_n^{conv}$ and $X_n^{targ}$ are the $n^{th}$ coefficient of the converted and target mel cepstra, $N$ is the dimension of MCC. The lower MCD indicates the smaller distortion. 

Note that MCD is an indirect measurement, which is not directly related to speech quality~\cite{machado2010voice}. As the durations of converted speech and target speech are different, we can not adopt perceptual evaluation of speech quality (PESQ)~\cite{rix2001perceptual} to measure the speech quality. The results of subjective evaluation reflect the actual perceptual quality of generated speech. 

Word error rate (WER) evaluated by automatic speech recognition (ASR) indicates intelligibility. The ASR model was Conformer~\cite{gulati2020conformer} based and trained on the 960h of LibriSpeech corpus.

\subsubsection{Subjective evaluation}

For subjective evaluation, we evaluated speech quality and speaker similarity using AB and ABX preference tests, respectively. For the AB test, A and B represent the randomly selected samples, where A and B have the same linguistic content. Listeners are asked to select a sample with better quality from A and B. For the ABX test, X refers to the reference sample of the target, A and B represent the converted samples randomly selected from the proposed and baseline method.  Then listeners are asked to choose the sample closer to the reference sample or no preference in terms of speaker similarity. 

We also assessed the speech naturalness with mean opinion score (MOS) test, where each listener is required to give opinion score on a five-point scale (5: excellent, 4: good, 3: fair, 2: poor, 1: bad). All the converted samples were used for listening tests. For each listener, we randomly selected 30 samples for listening test. Different listeners may listen to different samples. 20 listeners participated in all listening tests.

\section{Experimental results and analysis}
\label{sec:experiment results}
In this section, we report the experimental results to verify the effectiveness of proposed method under different test scenarios.

\subsection{Disentangling noise-invariant speaker and content representations}

\begin{figure}[!ht]
	\centering
	\centerline{\includegraphics[width=1.0\linewidth]{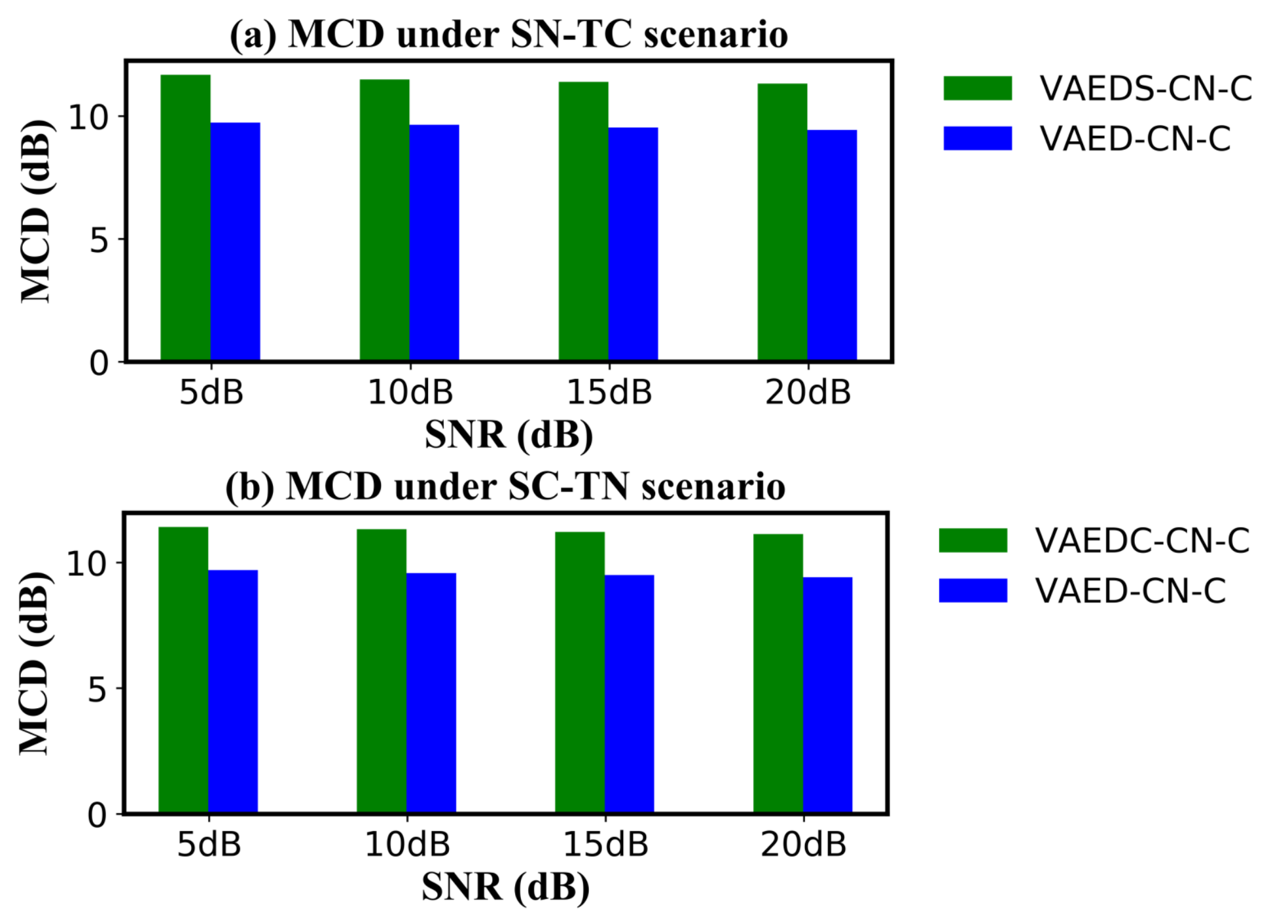}}
	\caption{Comparison of average MCD of proposed systems with and without domain adversarial training for the content encoder and speaker encoder under noisy scenarios.}
	\label{fig:noise-invariant}
\end{figure}

First, we compared the MCD scores of proposed systems under noisy scenarios to select best proposed system. Figure~\ref{fig:noise-invariant} (a) compares the average MCD between VAEDS-CN-C and VAED-CN-C under SN-TC scenario to verify that domain adversarial training for content encoder is necessary. VAED-CN-C outperforms VAEDS-CN-C significantly under different SNR levels  when source speech is corrupted by noises. Figure~\ref{fig:noise-invariant} (b) compares the average MCD between VAEDC-CN-C and VAED-CN-C under SC-TN scenario to verify that domain adversarial training for speaker encoder is necessary. VAED-CN-C outperforms VAEDC-CN-C significantly under different SNR levels when target speech is corrupted by noises. VAED-CN-C achieves the best performance among the proposed systems. This confirms that domain adversarial training for both content encoder and speaker encoder are necessary.

\begin{figure*}[!ht]
	\centering
	\centerline{\includegraphics[width=1.0\linewidth]{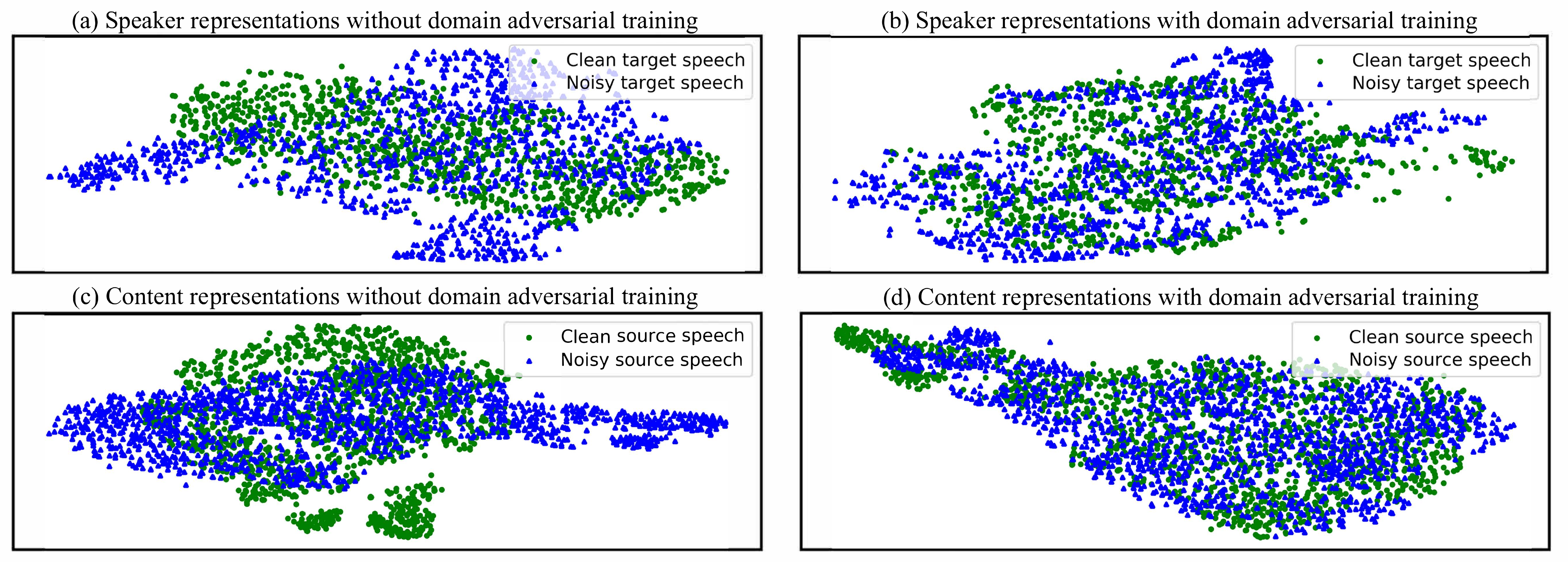}}
	\caption{An example of visualizing speaker representations and content representations from the male to female conversion pair under 5dB SNR of cafe noise, where the speaker representations and content representations are trained without and with domain adversarial training, respectively. (a) the speaker representations of VAEDC-CN-C are trained without DAT, (b) the speaker representations of VAED-CN-C are trained with DAT, (c) the content representations of VAEDS-CN-C are trained without DAT, (d) the content representations of VAED-CN-C are trained with DAT.}
	\label{fig:representations}
\end{figure*}

\begin{table*}[]
	\small
	\renewcommand\arraystretch{1.6}
	\centering
	\caption{A comparison of average MCD and WER between the proposed noise-robust voice conversion system and baseline systems under different noisy test scenarios. At run-time, SC-TN indicates that source speech is clean and target speech is noisy. SN-TC represents that source speech is noisy and target speech is clean. SN-TN indicates that both source and target speech are noisy.}
	\setlength{\tabcolsep}{0.9mm}{
\begin{tabular}{|c|c|cccc|cccc|cccc|cccc|}
	\hline
	\multirow{3}{*}{Scenario} & \multirow{3}{*}{System} & \multicolumn{4}{c|}{Street}                                                                  & \multicolumn{4}{c|}{Cafe}                                                                    & \multicolumn{4}{c|}{Babble}                                                                  & \multicolumn{4}{c|}{Hfchannel}                                                               \\ \cline{3-18} 
	&                         & \multicolumn{2}{c|}{MCD(dB)}                            & \multicolumn{2}{c|}{WER(\%)}       & \multicolumn{2}{c|}{MCD(dB)}                            & \multicolumn{2}{c|}{WER(\%)}       & \multicolumn{2}{c|}{MCD(dB)}                            & \multicolumn{2}{c|}{WER(\%)}       & \multicolumn{2}{c|}{MCD(dB)}                            & \multicolumn{2}{c|}{WER(\%)}       \\ \cline{3-18} 
	&                         & \multicolumn{1}{c|}{5dB}   & \multicolumn{1}{c|}{20dB}  & \multicolumn{1}{c|}{5dB}   & 20dB  & \multicolumn{1}{c|}{5dB}   & \multicolumn{1}{c|}{20dB}  & \multicolumn{1}{c|}{5dB}   & 20dB  & \multicolumn{1}{c|}{5dB}   & \multicolumn{1}{c|}{20dB}  & \multicolumn{1}{c|}{5dB}   & 20dB  & \multicolumn{1}{c|}{5dB}   & \multicolumn{1}{c|}{20dB}  & \multicolumn{1}{c|}{5dB}   & 20dB  \\ \hline
	\multirow{3}{*}{SC-TN}    & VAE-CN-C                & \multicolumn{1}{c|}{11.69} & \multicolumn{1}{c|}{11.44} & \multicolumn{1}{c|}{12.12} & 11.13 & \multicolumn{1}{c|}{11.56} & \multicolumn{1}{c|}{11.44} & \multicolumn{1}{c|}{12.71} & 11.63 & \multicolumn{1}{c|}{11.46} & \multicolumn{1}{c|}{11.37} & \multicolumn{1}{c|}{13.81} & 12.62 & \multicolumn{1}{c|}{11.52} & \multicolumn{1}{c|}{11.43} & \multicolumn{1}{c|}{13.35} & 12.14 \\ \cline{2-18} 
	& VAE-CD-C                & \multicolumn{1}{c|}{10.65} & \multicolumn{1}{c|}{10.53} & \multicolumn{1}{c|}{11.56} & 10.91 & \multicolumn{1}{c|}{10.67} & \multicolumn{1}{c|}{10.52} & \multicolumn{1}{c|}{11.97} & 11.13 & \multicolumn{1}{c|}{10.69} & \multicolumn{1}{c|}{10.56} & \multicolumn{1}{c|}{12.58} & 11.49 & \multicolumn{1}{c|}{10.69} & \multicolumn{1}{c|}{10.52} & \multicolumn{1}{c|}{12.25} & 11.37 \\ \cline{2-18} 
	& VAED-CN-C               & \multicolumn{1}{c|}{9.71}  & \multicolumn{1}{c|}{9.53}  & \multicolumn{1}{c|}{11.13} & 10.41 & \multicolumn{1}{c|}{9.66}  & \multicolumn{1}{c|}{9.53}  & \multicolumn{1}{c|}{11.29} & 10.36 & \multicolumn{1}{c|}{9.62}  & \multicolumn{1}{c|}{9.51}  & \multicolumn{1}{c|}{11.79} & 10.87 & \multicolumn{1}{c|}{9.61}  & \multicolumn{1}{c|}{9.47}  & \multicolumn{1}{c|}{11.49} & 10.53 \\ \hline
	\multirow{4}{*}{SN-TC}    & VAE-CN-C                & \multicolumn{1}{c|}{11.95} & \multicolumn{1}{c|}{11.59} & \multicolumn{1}{c|}{38.52} & 13.63 & \multicolumn{1}{c|}{11.98} & \multicolumn{1}{c|}{11.57} & \multicolumn{1}{c|}{44.56} & 14.56 & \multicolumn{1}{c|}{11.90} & \multicolumn{1}{c|}{11.58} & \multicolumn{1}{c|}{47.23} & 15.22 & \multicolumn{1}{c|}{11.82} & \multicolumn{1}{c|}{11.58} & \multicolumn{1}{c|}{48.58} & 16.18 \\ \cline{2-18} 
	& VAE-CD-C                & \multicolumn{1}{c|}{10.75} & \multicolumn{1}{c|}{10.62} & \multicolumn{1}{c|}{18.83} & 12.11 & \multicolumn{1}{c|}{10.87} & \multicolumn{1}{c|}{10.65} & \multicolumn{1}{c|}{22.35} & 12.16 & \multicolumn{1}{c|}{10.87} & \multicolumn{1}{c|}{10.64} & \multicolumn{1}{c|}{24.98} & 13.28 & \multicolumn{1}{c|}{10.94} & \multicolumn{1}{c|}{10.67} & \multicolumn{1}{c|}{25.28} & 14.34 \\ \cline{2-18} 
	& FHVAE-CN-CN             & \multicolumn{1}{c|}{11.40} & \multicolumn{1}{c|}{10.37} & \multicolumn{1}{c|}{22.43} & 12.50 & \multicolumn{1}{c|}{11.95} & \multicolumn{1}{c|}{10.83} & \multicolumn{1}{c|}{29.83} & 14.35 & \multicolumn{1}{c|}{10.90} & \multicolumn{1}{c|}{10.41} & \multicolumn{1}{c|}{40.15} & 13.30 & \multicolumn{1}{c|}{11.98} & \multicolumn{1}{c|}{10.90} & \multicolumn{1}{c|}{51.32} & 27.16 \\ \cline{2-18} 
	& VAED-CN-C               & \multicolumn{1}{c|}{9.75}  & \multicolumn{1}{c|}{9.42}  & \multicolumn{1}{c|}{19.95} & 12.43 & \multicolumn{1}{c|}{9.91}  & \multicolumn{1}{c|}{9.45}  & \multicolumn{1}{c|}{24.21} & 12.18 & \multicolumn{1}{c|}{9.68}  & \multicolumn{1}{c|}{9.47}  & \multicolumn{1}{c|}{26.54} & 13.62 & \multicolumn{1}{c|}{9.70}  & \multicolumn{1}{c|}{9.43}  & \multicolumn{1}{c|}{24.61} & 15.56 \\ \hline
	\multirow{3}{*}{SN-TN}    & VAE-CN-C                & \multicolumn{1}{c|}{12.06} & \multicolumn{1}{c|}{11.49} & \multicolumn{1}{c|}{40.45} & 14.52 & \multicolumn{1}{c|}{11.88} & \multicolumn{1}{c|}{11.47} & \multicolumn{1}{c|}{44.87} & 15.43 & \multicolumn{1}{c|}{11.84} & \multicolumn{1}{c|}{11.41} & \multicolumn{1}{c|}{47.45} & 15.18 & \multicolumn{1}{c|}{11.81} & \multicolumn{1}{c|}{11.48} & \multicolumn{1}{c|}{48.84} & 16.35 \\ \cline{2-18} 
	& VAE-CD-C                & \multicolumn{1}{c|}{10.82} & \multicolumn{1}{c|}{10.61} & \multicolumn{1}{c|}{18.91} & 12.71 & \multicolumn{1}{c|}{10.94} & \multicolumn{1}{c|}{10.67} & \multicolumn{1}{c|}{22.15} & 12.28 & \multicolumn{1}{c|}{10.95} & \multicolumn{1}{c|}{10.70} & \multicolumn{1}{c|}{25.21} & 13.21 & \multicolumn{1}{c|}{10.96} & \multicolumn{1}{c|}{10.74} & \multicolumn{1}{c|}{26.71} & 13.15 \\ \cline{2-18} 
	& VAED-CN-C               & \multicolumn{1}{c|}{9.88}  & \multicolumn{1}{c|}{9.50}  & \multicolumn{1}{c|}{19.79} & 12.81 & \multicolumn{1}{c|}{10.06} & \multicolumn{1}{c|}{9.51}  & \multicolumn{1}{c|}{24.39} & 12.33 & \multicolumn{1}{c|}{9.83}  & \multicolumn{1}{c|}{9.51}  & \multicolumn{1}{c|}{26.38} & 13.51 & \multicolumn{1}{c|}{9.82}  & \multicolumn{1}{c|}{9.45}  & \multicolumn{1}{c|}{25.21} & 15.25 \\ \hline
\end{tabular}}
	\label{table:all_systems}
\end{table*}

Then we show an example of speaker representations and content representations from the male to female conversion pair under 5dB SNR of cafe noise. The representations are projected to 2D using t-distributed stochastic neighbor embedding (t-SNE)~\cite{2008Visualizing}. Figure~\ref{fig:representations} (a) shows the speaker representations extracted by VAEDC-CN-C from clean and noisy target speech. The speaker encoder of VAEDC-CN-C takes clean or noisy speech as input, and the output of decoder is always the clean speech. Hence the speaker encoder learns to project the speaker representations extracted from clean speech and noisy speech into the same subspace. As a result, the speaker representations from two domains are partially overlapped. Due to the limited capability of projecting the speaker representations from different domains into the same space, the speaker representations of clean speech and noisy speech from the same speaker have different distributions when speaker encoder is trained without domain adversarial training. Figure~\ref{fig:representations} (b) shows that the speaker representations extracted by VAED-CN-C have similar distributions with the help of DAT for speaker encoder. Figure~\ref{fig:representations} (c) shows that the content representations  extracted by VAEDS-CN-C from clean and noisy source speech belong to different distributions. With the help of domain adversarial training for the content encoder of VAED-CN-C, the content representations are overlapped shown in Figure~\ref{fig:representations} (d). It was observed that the proposed system VAED-CN-C successfully learns to extract noise-invariant speaker and content representations with domain adversarial training. Note that we also observed some mis-alignments on the contours of distributions of extracted content and speaker representations from VAED-CN-C, respectively. This indicates that there is still room for improvement.

With the empirical observations above, we use VAED-CN-C in the rest of the experiments and compare it with baseline systems.

\subsection{Objective evaluation}
The objective evaluation results of different noise-robust voice conversion methods under three noisy test scenarios are shown in Table~\ref{table:all_systems}. Due to limited space, we report MCD and WER results at different SNR levels of two seen noise types: street and cafe, two unseen noise types: \emph{babble and hfchannel}, respectively.

Under the SC-TN noisy scenario, noise is only from target speech, which affects the learned speaker representations. We first evaluated system performance under seen noise types. As the SNR increases from 5dB to 20dB, the MCD and WER scores of all systems decrease for street and cafe noises. Under the same noise type and SNR level, as the speaker representations extracted from clean and noisy speech have different distributions while the output of decoder is clean speech, this mismatch degrades the performance of decoder, VAE-CN-C gets the highest MCD and WER scores, which indicate that VAE-CN-C performs worst among all the systems. By using speech enhancement model DCCRN as a pre-processing module to get de-noised speech, VAE-CD-C performs consistently better than VAE-CN-C. By learning noise-invariant speaker representations which reduce the gap in distributions of different domains, our proposed system VAED-CN-C consistently outperforms the baseline systems at different SNR levels of street noise and cafe noise. According to WER, VAED-CN-C consistently performs better than VAE-CN-C and VAE-CD-C. Then we evaluated system performance under unseen noise types. It was observed that as the SNR level increases, the MCD and WER scores decrease for all systems. Additionally, VAED-CN-C consistently outperforms the baseline systems at all evaluated SNR levels in terms of MCD and WER. Finally, we compared the system performance between seen and unseen noise types. It was observed that VAE-CN-C and VAE-CD-C perform worse at all evaluated SNR levels of unseen noise types than that of seen noise types in terms of WER. Our proposed VAED-CN-C performs more stable between seen and unseen noise types.

Under the SN-TC noisy scenario, noise is only from source speech, which affects the learned content representations. We first evaluated system performance under seen noise types. When the SNR level increases from 5dB to 20dB, the MCD and WER scores of all systems decrease under street and cafe noisy conditions. Compared at the same SNR level, VAE-CN-C performs worst under different SNR levels among all the systems, because the content representations extracted from clean and noisy domains have different distributions, which degrade the performance of decoder. VAE-CD-C outperforms VAE-CN-C at all evaluated SNR levels due to taking denoised speech as input. Since the content representation of FHVAE is noise-invariant~\cite{hsu2017unsupervised}, FHVAE-CN-CN performs better than VAE-CN-C. As speech enhancement model DCCRN introduces some processing artifacts, while domain adversarial training projects the representations from clean and noisy domains into the same subspace which mitigates the deterioration, our proposed system VAED-CN-C is able to perform better than VAE-CD-C in terms of MCD. According to WER, VAE-CD-C performs best. VAED-CN-C is slightly worse than VAE-CD-C under different SNR levels, since there still exists mis-alignments on the contours of distributions of extracted content reresentations shown in Figure~\ref{fig:representations}. Then we evaluated system performance under unseen noise types. It was observed that VAED-CN-C consistently outperforms VAE-CN-C, VAE-CD-C and FHVAE-CN-CN at different SNR levels in terms of MCD. According to WER, VAE-CD-C performs best. VAED-CN-C outperforms VAE-CN-C and FHVAE-CN-CN significantly. Finally, we compared the system performance between seen and unseen noise types. VAE-CN-C, VAE-CD-C and VAED-CN-C perform similarly between seen and unseen noise types in terms of MCD, while FHVAE-CN-CN performs worse under hfchannel noisy condition. All the systems perform  worse under unseen noise types than that of seen noise types according to WER.

Under the SN-TN noisy scenario, noise is from both source speech and target speech, which affects the learned content representations and speaker representations at the same time. First, we compared our proposed system with baseline systems. At all evaluated SNR levels of seen and unseen noise types, it was observed that VAED-CN-C consistently outperforms VAE-CN-C and VAE-CD-C in terms of MCD. According to WER, VAED-CN-C performs better than VAE-CN-C, while performing slightly worse than VAE-CD-C. Then we compared the system performance between seen and unseen noise types.  The performances between seen and unseen noise types of all systems are similar in terms of MCD. Additionally, all the systems perform worse under unseen noise types than that of seen noise types according to WER.

We also compared all the systems under clean scenario where both source speech and target speech are clean. The MCD and WER scores are shown in Table~\ref{table:sc-tc}. As the decoder of VAE-C-C only takes representation from clean speech as input, while the decoder of VAE-CN-C takes representations from clean or noisy speech as input which have differerent distributions, 
the performance of VAE-CN-C seriously degrades compared with VAE-C-C. By obtaining de-noised speech, VAE-CD-C performs better than VAE-CN-C. As FHVAE-CN-CN is able to disentangle noise from content representation, it performs better than VAE-CN-C. Our proposed method VAED-CN-C makes speaker and content representations extracted from noisy and clean speech indistinguishable, VAED-CN-C achieves better performance in terms of MCD. According to WER, VAED-CN-C outperforms VAE-CN-C and VAE-CD-C, while performing worse than FHVAE-CN-CN.

\begin{table}[]
	\small
	\renewcommand\arraystretch{1.5}
	\centering
	\caption{A comparison of MCD and WER between proposed system and baseline systems under clean scenario. At run-time, SC-TC indicates that both source speech and target speech are clean.}
	\setlength{\tabcolsep}{3.8mm}{
		\begin{tabular}{|c|c|c|c|}
			\hline
			Scenario               & System    & MCD(dB) & WER(\%) \\ \hline
			\multirow{5}{*}{SC-TC} & VAE-C-C   & 9.39   & 8.93    \\ \cline{2-4} 
			& VAE-CN-C  & 11.32   & 10.74   \\ \cline{2-4} 
			& VAE-CD-C  & 10.45   & 9.81    \\ \cline{2-4} 
			& FHVAE-CN-CN & 10.34   & 8.62    \\ \cline{2-4} 
			& VAED-CN-C & 9.47   & 9.25    \\ \hline
	\end{tabular}}
	\label{table:sc-tc}
\end{table}

\begin{figure}[!ht]
	\centering
	\centerline{\includegraphics[width=1.0\linewidth]{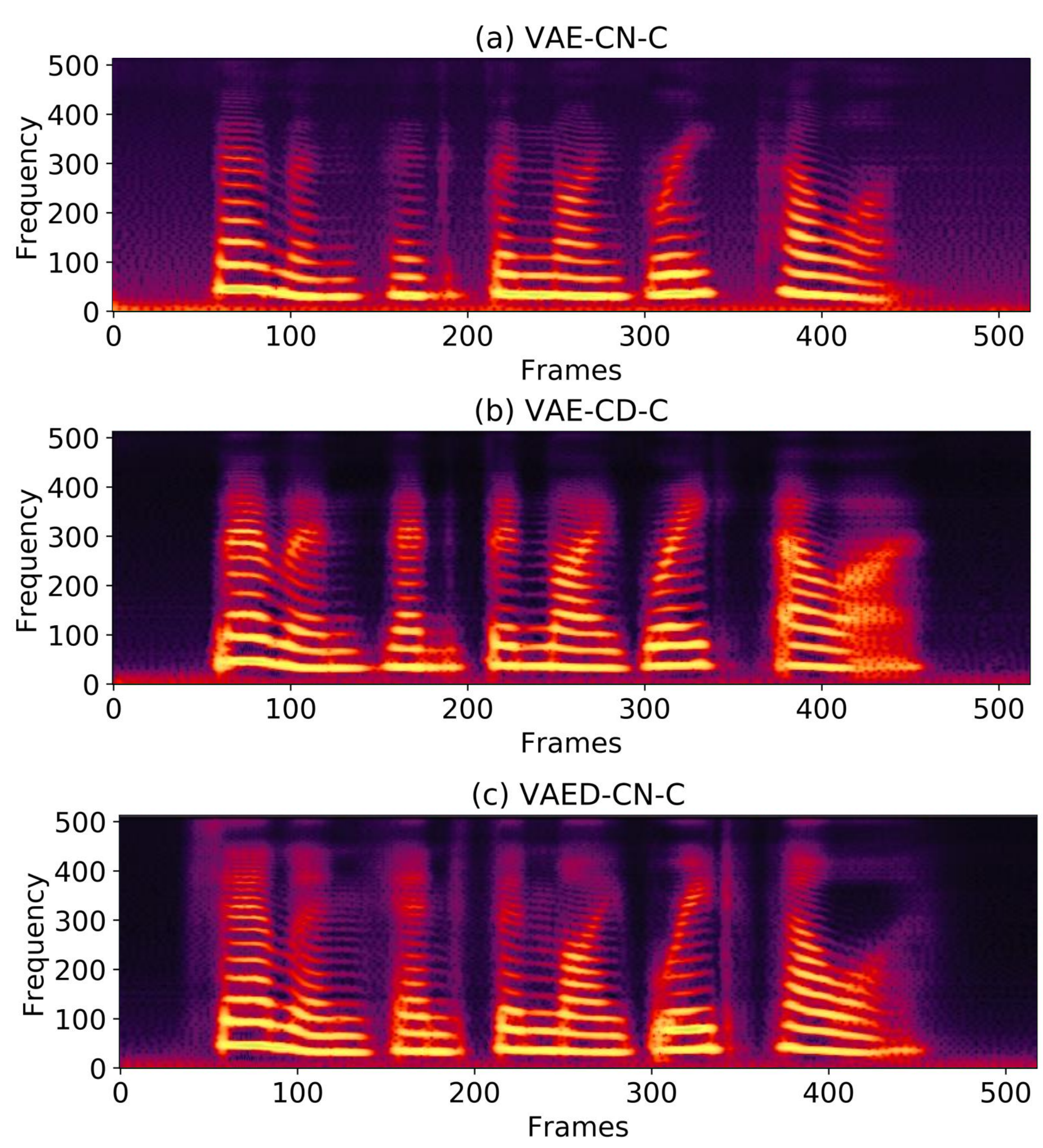}}
	\caption{Spectrogram for a sentence ``Famine had been my great ally", converted from male speaker to female  speaker under SN-TN scenario with 5dB SNR of cafe noise, and generated by three different systems (a) VAE-CN-C, (b) VAE-CD-C, (c) VAED-CN-C.}
	\label{fig:example}
\end{figure}

Finally, Figure~\ref{fig:example} shows the spectrogram of the converted speech from  male speaker to female speaker by the systems VAE-CN-C, VAE-CD-C, VAED-CN-C under SN-TN scenario with 5dB SNR of cafe noise,  respectively. Figure~\ref{fig:example} (a) shows that the converted speech by VAE-CN-C is corrupted by noise, and high-frequency part is lost. Figure~\ref{fig:example} (b) suggests that the converted speech from VAE-CD-C is clean. However, DCCRN introduces extra distortion. Figure~\ref{fig:example} (c) shows that our proposed method effectively reduces the noise from the noisy input speech while the speech distortion is minimized.

In summary, learning noise-invariant representations is effective to predict clean converted spectrum under different noise types and SNR levels.

\subsection{Subjective evaluation}

\begin{figure}[!ht]
	\centering
	\centerline{\includegraphics[width=1.0\linewidth]{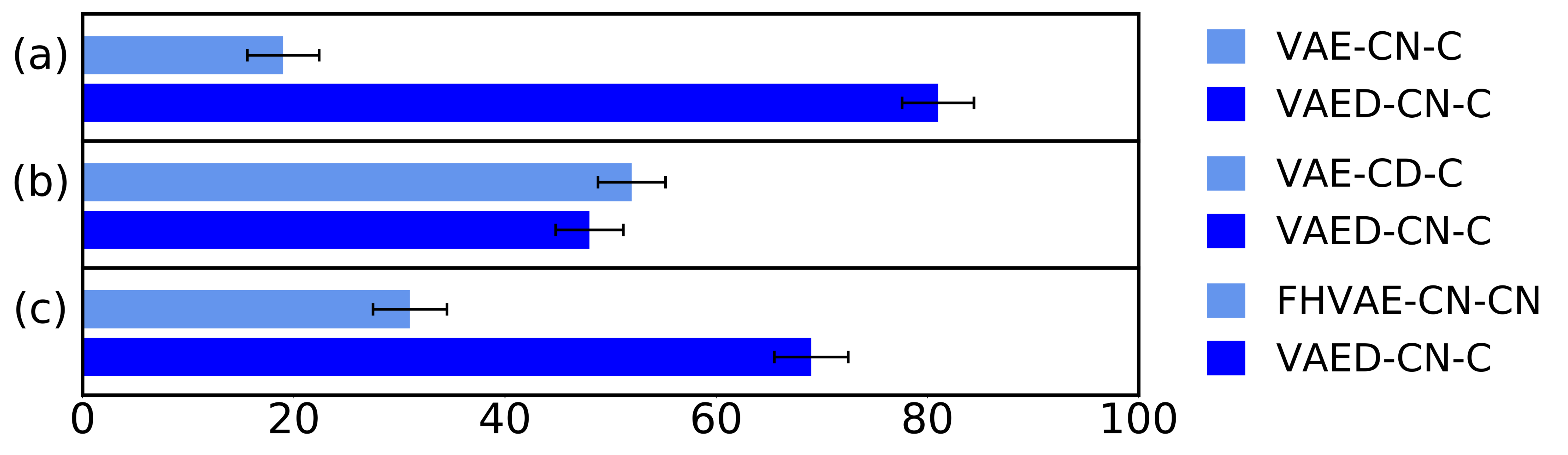}}
	\caption{Speech quality preference tests under 5dB SNR of six  types of noise with 95\% confidence intervals for (a) VAE-CN-C vs VAED-CN-C, (b) VAE-CD-C vs VAED-CN-C, (c) FHVAE-CN-CN vs VAED-CN-C.}
	\label{fig:ab-5dB}
\end{figure}

Figure~\ref{fig:ab-5dB} reports the AB listening tests for speech quality under six types of noise with 5dB SNR. First, we would like to confirm that domain adversarial training helps to improve speech quality under noisy scenarios with low SNR level. Figure~\ref{fig:ab-5dB} (a) suggests that our proposed VAED-CN-C significantly outperforms VAE-CN-C under 5dB SNR. Then we compared VAED-CN-C with noise-robust voice conversion baseline systems. Figure~\ref{fig:ab-5dB} (b) shows that the preference scores of VAED-CN-C and VAE-CD-C fall into each other's confidence intervals,
which means they are not significantly different. Figure~\ref{fig:ab-5dB} (c) shows that VAED-CN-C clearly outperforms FHVAE-CN-CN.

\begin{figure}[!ht]
	\centering
	\centerline{\includegraphics[width=1.0\linewidth]{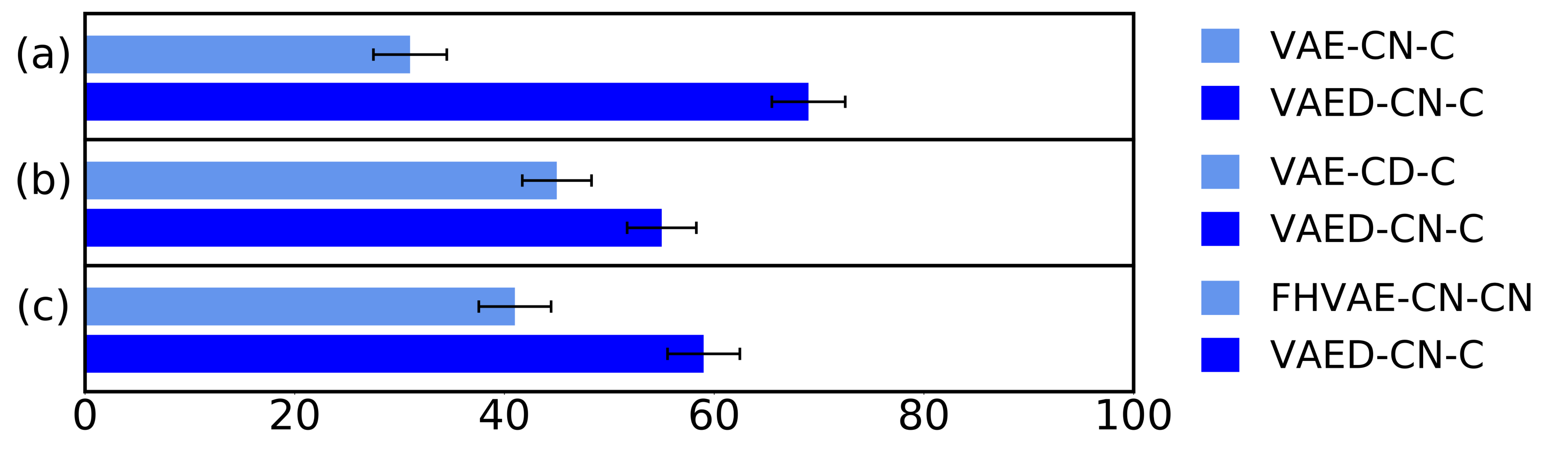}}
	\caption{Speech quality preference tests under 20dB SNR of six  types of noise with 95\% confidence intervals for (a) VAE-CN-C  vs VAED-CN-C, (b) VAE-CD-C vs VAED-CN-C, (c) FHVAE-CN-CN vs VAED-CN-C.}
	\label{fig:ab-20dB}
\end{figure}

Figure~\ref{fig:ab-20dB} reports the AB listening tests for speech quality under six types of noise with 20dB SNR. First, we would like to confirm that domain adversarial training helps to improve speech quality under noisy scenarios with high SNR level. Figure~\ref{fig:ab-20dB} (a) suggests that our proposed VAED-CN-C significantly outperforms VAE-CN-C under 20dB SNR. Then we compared VAED-CN-C with other noise-robust voice conversion systems. Figure~\ref{fig:ab-20dB} (b) and (c) show that VAED-CN-C outperforms VAE-CD-C and FHVAE-CN-CN in terms of speech quality. 

\begin{figure}[!ht]
	\centering
	\centerline{\includegraphics[width=1.0\linewidth]{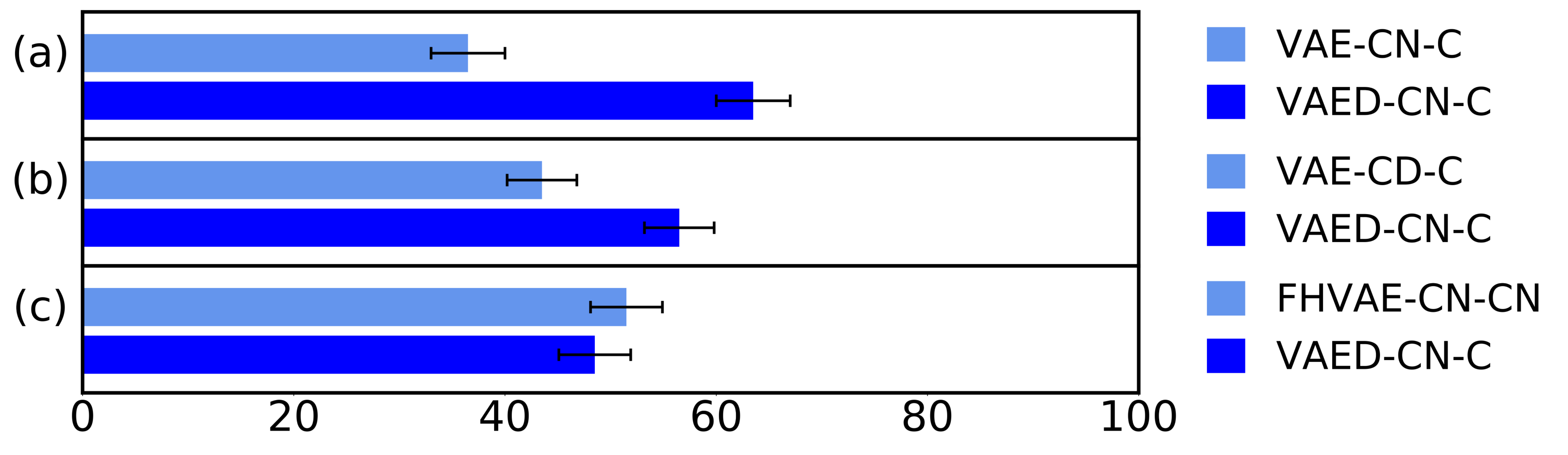}}
	\caption{Speech quality preference tests under clean scenario with 95\% confidence intervals for (a) VAE-CN-C vs VAED-CN-C, (b) VAE-CD-C vs VAED-CN-C, (c) FHVAE-CN-CN vs VAED-CN-C.}
	\label{fig:ab-clean}
\end{figure}

Figure~\ref{fig:ab-clean} reports the AB listening tests for speech quality under clean scenario. Figure~\ref{fig:ab-clean} (a) and (b) suggest that VAED-CN-C outperforms VAE-CN-C and VAE-CD-C. Figure~\ref{fig:ab-clean} (c) shows that FHVAE-CN-CN performs better than VAED-CN-C. 

\begin{figure}[!ht]
	\centering
	\centerline{\includegraphics[width=1.0\linewidth]{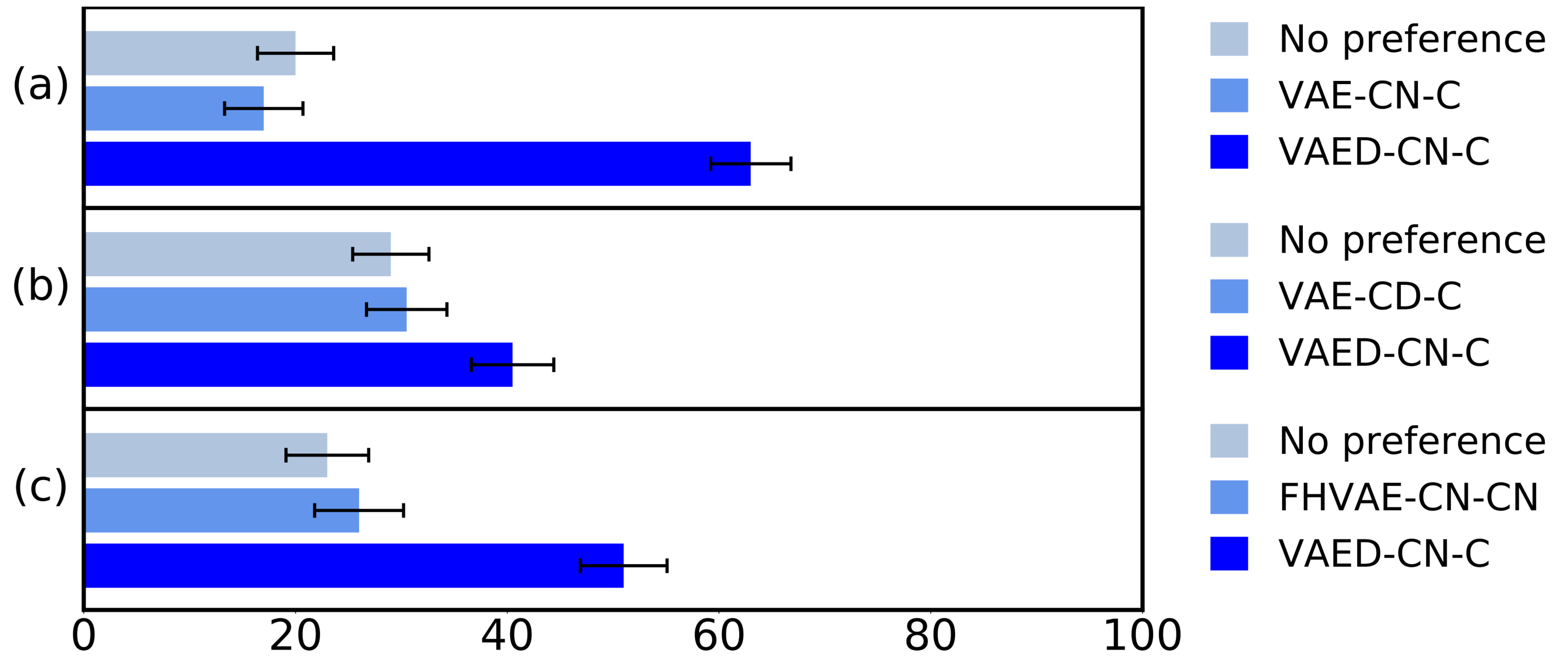}}
	\caption{Similarity preference tests under 5dB SNR  of six  types of noise with 95\% confidence intervals for (a) VAE-CN-C  vs VAED-CN-C, (b) VAE-CD-C vs VAED-CN-C, (c) FHVAE-CN-CN vs VAED-CN-C.}
	\label{fig:abx-5dB}
\end{figure}

The speaker similarity ABX tests under six types of noise with 5dB SNR are presented in Figure~\ref{fig:abx-5dB}. First, we would like to confirm that domain adversarial training helps to improve speaker similarity under noisy scenario with low SNR level. Figure~\ref{fig:abx-5dB} (a) suggests our proposed VAED-CN-C significantly outperforms VAE-CN-C. Then we compared VAED-CN-C with other noise-robust voice conversion systems. Figure~\ref{fig:abx-5dB} (b) and (c) show that VAED-CN-C outperforms VAE-CD-C and FHVAE-CN-CN, respectively.

\begin{figure}[!ht]
	\centering
	\centerline{\includegraphics[width=1.0\linewidth]{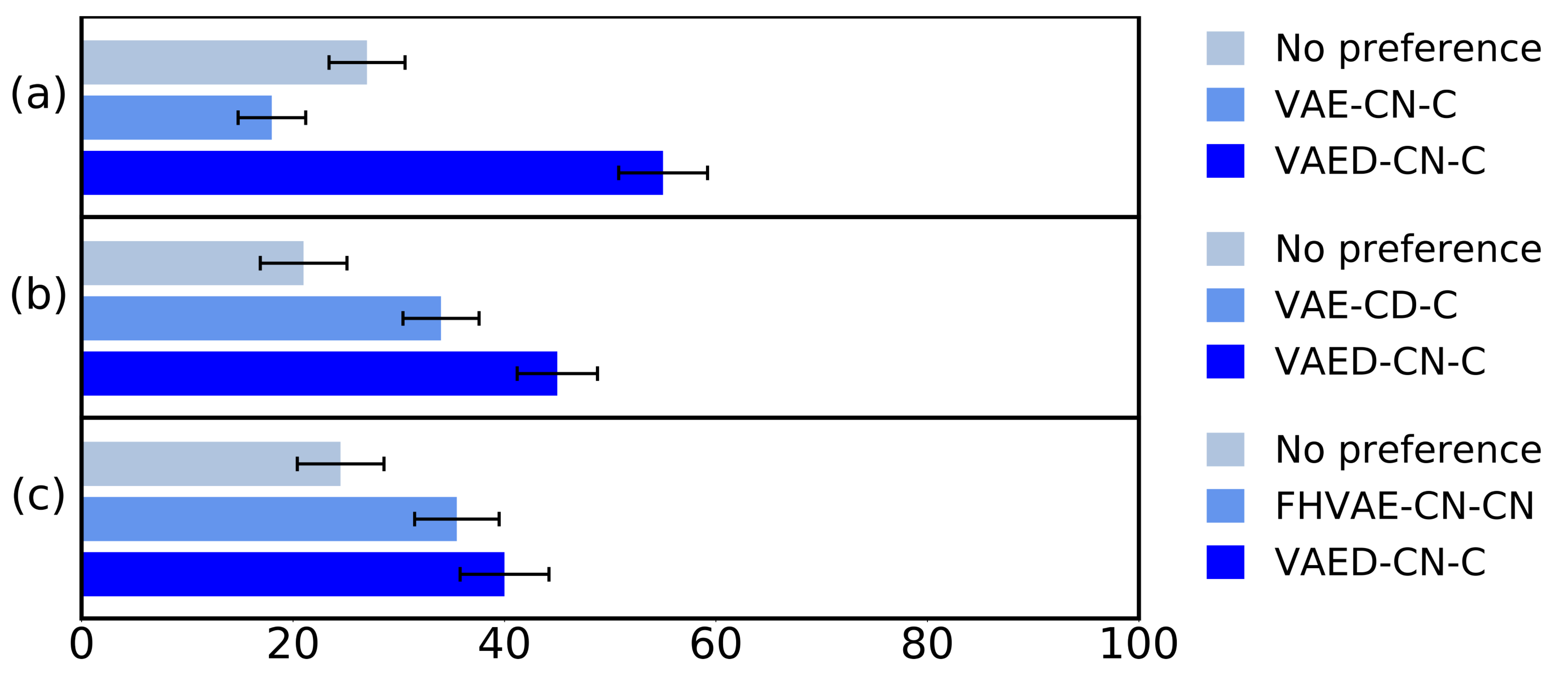}}
	\caption{Similarity preference tests under 20dB SNR of six noise types with 95\% confidence intervals for (a) VAE-CN-C  vs VAED-CN-C, (b) VAE-CD-C vs VAED-CN-C, (c) FHVAE-CN-CN vs VAED-CN-C.}
	\label{fig:abx-20dB}
\end{figure}

The speaker similarity ABX tests under six types of noise with 20dB SNR are presented in Figure~\ref{fig:abx-20dB}. Figure~\ref{fig:abx-20dB} (a) suggests our proposed VAED-CN-C significantly outperforms VAE-CN-C under 20dB SNR level. Comparing with other noise-robust voice conversion systems,  Figure~\ref{fig:abx-20dB} (b) and (c) show that VAED-CN-C outperforms VAE-CD-C and FHVAE-CN-CN, respectively.

\begin{figure}[!ht]
	\centering
	\centerline{\includegraphics[width=1.0\linewidth]{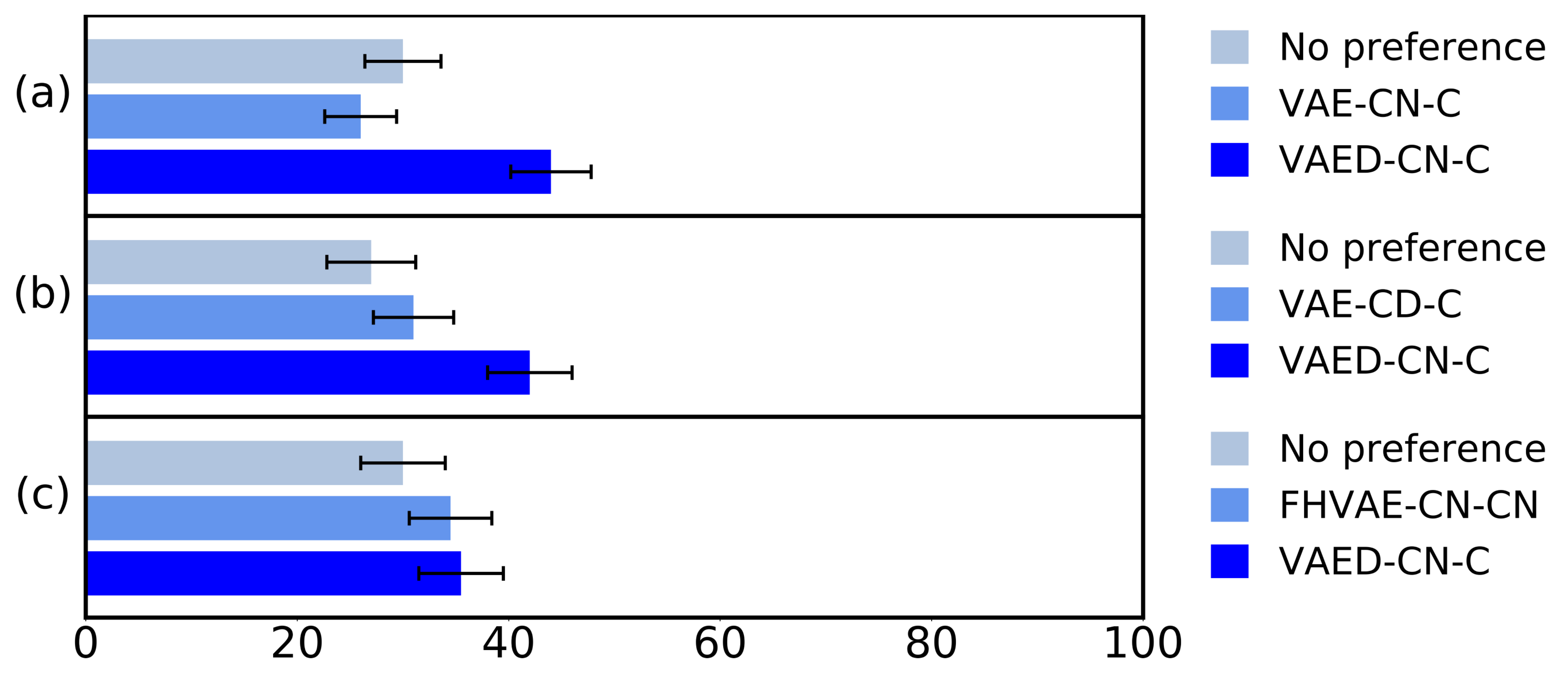}}
	\caption{Similarity preference tests under clean scenario with 95\% confidence intervals for (a) VAE-CN-C vs VAED-CN-C, (b) VAE-CD-C vs VAED-CN-C, (c) FHVAE-CN-CN vs VAED-CN-C.}
	\label{fig:abx-clean}
\end{figure}

The speaker similarity ABX tests under clean scenario are presented in Figure~\ref{fig:abx-clean}. Figure~\ref{fig:abx-clean} (a) and (b) suggest our proposed VAED-CN-C clearly outperforms VAE-CN-C and VAE-CD-C. Figure~\ref{fig:abx-clean} (c) shows that FHVAE-CN-CN performs slightly worse than VAED-CN-C.

\begin{figure}[!ht]
	\centering
	\centerline{\includegraphics[width=1.0\linewidth]{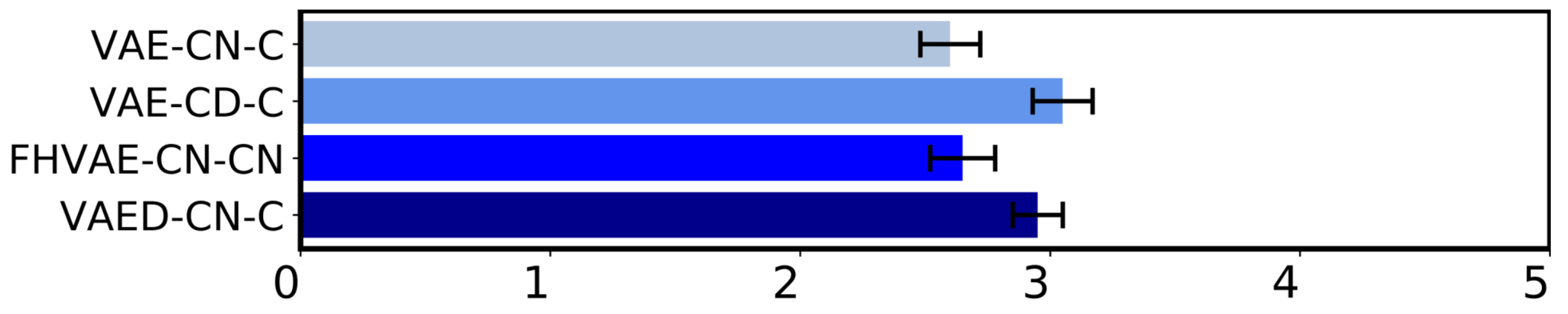}}
	\caption{Mean opinion score listening tests among different systems under  six types of noise with 5dB SNR. Error bar represents 95\% confidence interval.}
	\label{fig:mod-5dB}
\end{figure}

\begin{figure}[!ht]
	\centering
	\centerline{\includegraphics[width=1.0\linewidth]{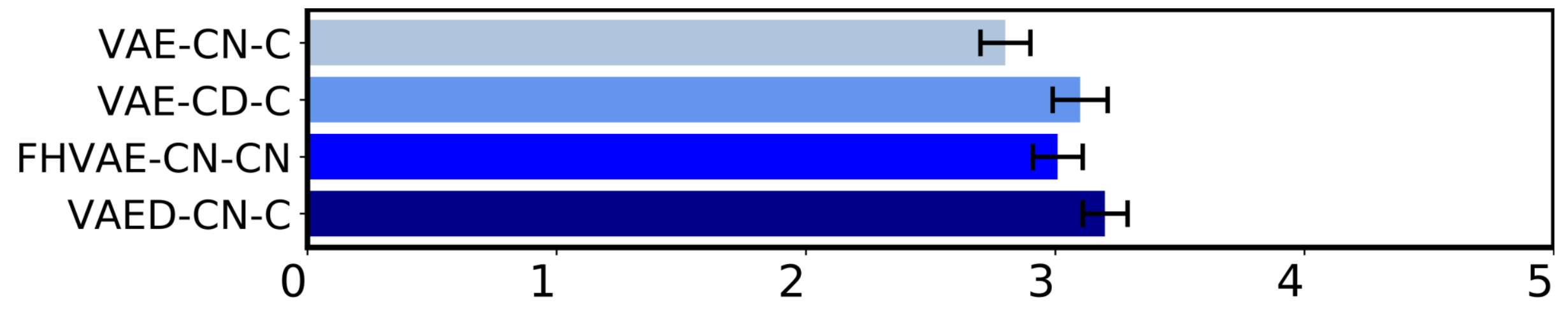}}
	\caption{Mean opinion score listening tests among different systems under  six types of noise with 20dB SNR. Error bar represents 95\% confidence interval.}
	\label{fig:mod-20dB}
\end{figure}

\begin{figure}[!ht]
	\centering
	\centerline{\includegraphics[width=1.0\linewidth]{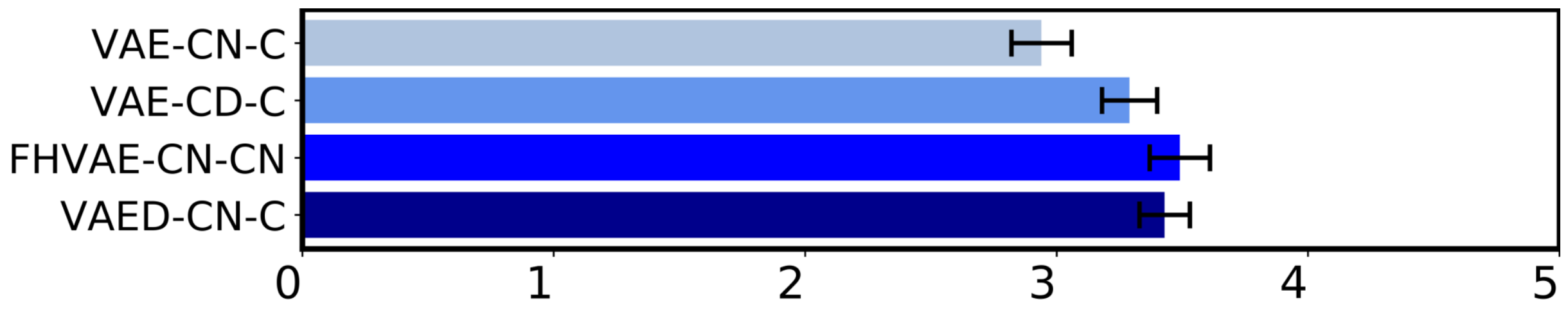}}
	\caption{Mean opinion score listening tests among different systems under clean scenario. Error bar represents 95\% confidence interval.}
	\label{fig:mod-clean}
\end{figure}

We evaluated the naturalness of speech under noisy scenario with 5dB SNR, 20dB SNR, and clean scenario through MOS tests, respectively. Figure~\ref{fig:mod-5dB} shows the MOS results under noisy scenario with 5dB SNR. It suggests that VAE-CD-C, FHVAE-CN-CN, and VAED-CN-C outperform VAE-CN-C.  Furthermore, VAED-CN-C performs slightly worse than VAE-CD-C and better than FHVAE-CN-CN.  Figure~\ref{fig:mod-20dB} shows the MOS results under noisy scenario with 20dB SNR.  VAE-CD-C, FHVAE-CN-CN, and VAED-CN-C consistently outperform VAE-CN-C, and our proposed VAED-CN-C achieves the best MOS score. Figure~\ref{fig:mod-clean} shows the MOS results under clean scenario. We observed that VAED-CN-C performs better than VAE-CD-C while slightly worse than FHVAE-CN-CN. The synthesized samples can be found on the website~\footnote{https://dhqadg.github.io/noise-robust/}.

\begin{figure}[!ht]
	\centering
	\centerline{\includegraphics[width=1.0\linewidth]{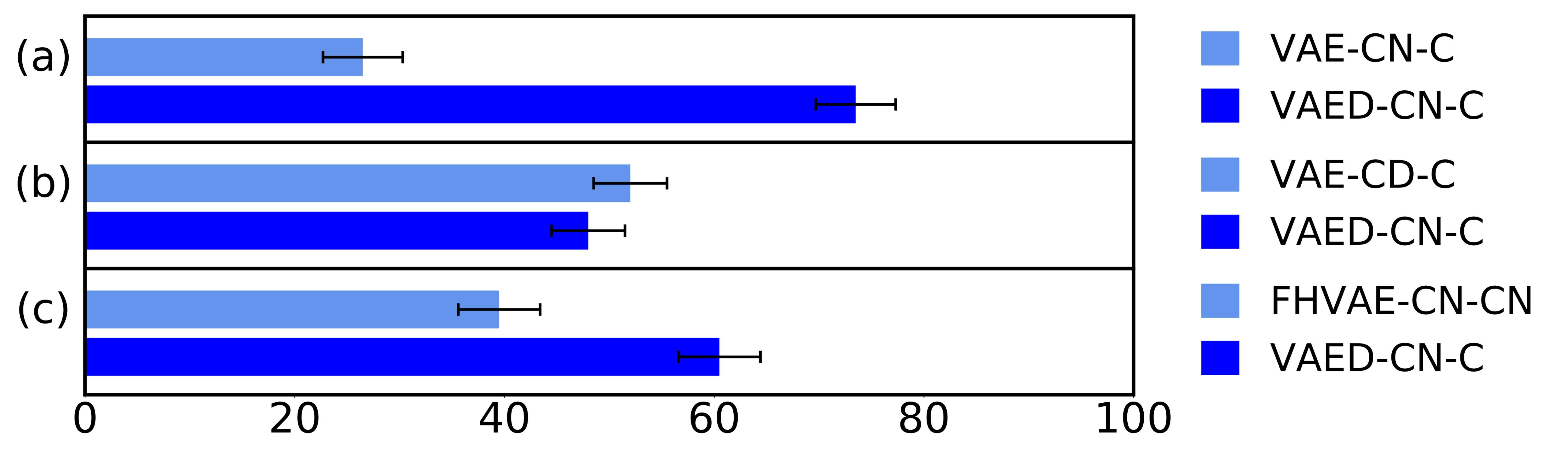}}
	\caption{Speech quality preference tests under real noisy  scenario with 95\% confidence intervals for (a) VAE-CN-C vs VAED-CN-C, (b) VAE-CD-C vs VAED-CN-C, (c) FHVAE-CN-CN vs VAED-CN-C.}
	\label{fig:ab-real}
\end{figure}

Finally, Figure~\ref{fig:ab-real} reports the AB listening tests for speech quality under real noisy scenario. Figure~\ref{fig:ab-real} (a) validates that our proposed VAED-CN-C significantly outperforms VAE-CN-C.  Figure~\ref{fig:ab-real} (b) shows VAED-CN-C performs slightly worse than VAE-CD-C. Figure~\ref{fig:ab-real} (c) shows that VAED-CN-C outperforms FHVAE-CN-CN.

\section{Conclusion and future work}
\label{sec:conclusion}
In this paper, we propose a novel noise-robust voice conversion framework. This framework can synthesize clean converted speech under complex noisy conditions, where both source speech and target speech at run-time can be corrupted by seen and unseen noise types. Specifically, based on the framework of encoder-decoder, we integrate disentangling speaker and content representations technique with domain adversarial training technique. Domain adversarial training makes speaker representations and content representations extracted by the speaker encoder and content encoder from clean speech and noisy speech in the same space, which are noise-invariant. Therefore, the noise-invariant representations can be taken as input for the decoder to predict clean converted spectrum. The experimental results demonstrate that our proposed method can synthesize clean converted speech under complex noisy conditions. The speech quality and speaker similarity are improved compared with baseline systems. In the future, we will continue to focus on the neural network architecture to make it work better under noisy conditions with low SNR level.

\section{Acknowledgements}

This work was supported by the National Key Research and Development Program of China (No. 2020AAA0108600).

This work was also supported by the National Research Foundation, Singapore under its AI Singapore Programme (AISG Award No: AISG-GC-2019-002) and (AISG Award No: AISG-100E-2018-006), and its National Robotics Programme (Grant No. 192 25 00054), and in part by RIE2020 Advanced Manufacturing and Engineering Programmatic Grants A1687b0033, and A18A2b0046.


\bibliography{mybibfile}

\begin{thebibliography}{62}
\providecommand{\natexlab}[1]{#1}
\providecommand{\url}[1]{\texttt{#1}}
\providecommand{\urlprefix}{URL }
\expandafter\ifx\csname urlstyle\endcsname\relax
  \providecommand{\doi}[1]{doi:\discretionary{}{}{}#1}\else
  \providecommand{\doi}[1]{doi:\discretionary{}{}{}\begingroup
  \urlstyle{rm}\url{#1}\endgroup}\fi
\providecommand{\bibinfo}[2]{#2}

\bibitem[{Mohammadi and Kain(2017)}]{mohammadi2017overview}
\bibinfo{author}{S.~H. Mohammadi}, \bibinfo{author}{A.~Kain},
  \bibinfo{title}{An overview of voice conversion systems},
  \bibinfo{journal}{Speech Communication} \bibinfo{volume}{88}
  (\bibinfo{year}{2017}) \bibinfo{pages}{65--82}.

\bibitem[{Mouchtaris \emph{et~al.}(2004)Mouchtaris, Van~der Spiegel, and
  Mueller}]{mouchtaris2004spectral}
\bibinfo{author}{A.~Mouchtaris}, \bibinfo{author}{J.~Van~der Spiegel},
  \bibinfo{author}{P.~Mueller}, \bibinfo{title}{A spectral conversion approach
  to the iterative Wiener filter for speech enhancement}, in:
  \bibinfo{booktitle}{IEEE international conference on multimedia and expo
  (ICME)}, vol.~\bibinfo{volume}{3}, \bibinfo{organization}{IEEE},
  \bibinfo{pages}{1971--1974}, \bibinfo{year}{2004}.

\bibitem[{Benisty and Malah(2011)}]{benisty2011voice}
\bibinfo{author}{H.~Benisty}, \bibinfo{author}{D.~Malah}, \bibinfo{title}{Voice
  conversion using GMM with enhanced global variance}, in:
  \bibinfo{booktitle}{Twelfth Annual Conference of the International Speech
  Communication Association}, \bibinfo{year}{2011}.

\bibitem[{Stylianou \emph{et~al.}(1998)Stylianou, Capp{\'e}, and
  Moulines}]{stylianou1998continuous}
\bibinfo{author}{Y.~Stylianou}, \bibinfo{author}{O.~Capp{\'e}},
  \bibinfo{author}{E.~Moulines}, \bibinfo{title}{Continuous probabilistic
  transform for voice conversion}, \bibinfo{journal}{IEEE Transactions on
  speech and audio processing} \bibinfo{volume}{6}~(\bibinfo{number}{2})
  (\bibinfo{year}{1998}) \bibinfo{pages}{131--142}.

\bibitem[{Toda \emph{et~al.}(2007)Toda, Black, and Tokuda}]{toda2007voice}
\bibinfo{author}{T.~Toda}, \bibinfo{author}{A.~W. Black},
  \bibinfo{author}{K.~Tokuda}, \bibinfo{title}{Voice conversion based on
  maximum-likelihood estimation of spectral parameter trajectory},
  \bibinfo{journal}{IEEE Transactions on Audio, Speech, and Language
  Processing} \bibinfo{volume}{15}~(\bibinfo{number}{8}) (\bibinfo{year}{2007})
  \bibinfo{pages}{2222--2235}.

\bibitem[{Erro \emph{et~al.}(2009)Erro, Moreno, and Bonafonte}]{erro2009voice}
\bibinfo{author}{D.~Erro}, \bibinfo{author}{A.~Moreno},
  \bibinfo{author}{A.~Bonafonte}, \bibinfo{title}{Voice conversion based on
  weighted frequency warping}, \bibinfo{journal}{IEEE Transactions on Audio,
  Speech, and Language Processing} \bibinfo{volume}{18}~(\bibinfo{number}{5})
  (\bibinfo{year}{2009}) \bibinfo{pages}{922--931}.

\bibitem[{Godoy \emph{et~al.}(2011)Godoy, Rosec, and Chonavel}]{godoy2011voice}
\bibinfo{author}{E.~Godoy}, \bibinfo{author}{O.~Rosec},
  \bibinfo{author}{T.~Chonavel}, \bibinfo{title}{Voice conversion using dynamic
  frequency warping with amplitude scaling, for parallel or nonparallel
  corpora}, \bibinfo{journal}{IEEE Transactions on Audio, Speech, and Language
  Processing} \bibinfo{volume}{20}~(\bibinfo{number}{4}) (\bibinfo{year}{2011})
  \bibinfo{pages}{1313--1323}.

\bibitem[{Tian \emph{et~al.}(2015)Tian, Wu, Lee, Hy, Chng, and
  Dong}]{tian2015sparse}
\bibinfo{author}{X.~Tian}, \bibinfo{author}{Z.~Wu}, \bibinfo{author}{S.~W.
  Lee}, \bibinfo{author}{N.~Q. Hy}, \bibinfo{author}{E.~S. Chng},
  \bibinfo{author}{M.~Dong}, \bibinfo{title}{Sparse representation for
  frequency warping based voice conversion}, in: \bibinfo{booktitle}{2015 IEEE
  International Conference on Acoustics, Speech and Signal Processing
  (ICASSP)}, \bibinfo{organization}{IEEE}, \bibinfo{pages}{4235--4239},
  \bibinfo{year}{2015}.

\bibitem[{Takashima \emph{et~al.}(2012)Takashima, Takiguchi, and
  Ariki}]{takashima2012exemplar}
\bibinfo{author}{R.~Takashima}, \bibinfo{author}{T.~Takiguchi},
  \bibinfo{author}{Y.~Ariki}, \bibinfo{title}{Exemplar-based voice conversion
  in noisy environment}, in: \bibinfo{booktitle}{2012 IEEE Spoken Language
  Technology Workshop (SLT)}, \bibinfo{organization}{IEEE},
  \bibinfo{pages}{313--317}, \bibinfo{year}{2012}.

\bibitem[{Wu \emph{et~al.}(2014)Wu, Virtanen, Chng, and Li}]{wu2014exemplar}
\bibinfo{author}{Z.~Wu}, \bibinfo{author}{T.~Virtanen}, \bibinfo{author}{E.~S.
  Chng}, \bibinfo{author}{H.~Li}, \bibinfo{title}{Exemplar-based sparse
  representation with residual compensation for voice conversion},
  \bibinfo{journal}{IEEE/ACM Transactions on Audio, Speech, and Language
  Processing} \bibinfo{volume}{22}~(\bibinfo{number}{10})
  (\bibinfo{year}{2014}) \bibinfo{pages}{1506--1521}.

\bibitem[{Tian \emph{et~al.}(2017)Tian, Lee, Wu, Chng, and
  Li}]{tian2017exemplar}
\bibinfo{author}{X.~Tian}, \bibinfo{author}{S.~W. Lee},
  \bibinfo{author}{Z.~Wu}, \bibinfo{author}{E.~S. Chng},
  \bibinfo{author}{H.~Li}, \bibinfo{title}{An exemplar-based approach to
  frequency warping for voice conversion}, \bibinfo{journal}{IEEE/ACM
  Transactions on Audio, Speech, and Language Processing}
  \bibinfo{volume}{25}~(\bibinfo{number}{10}) (\bibinfo{year}{2017})
  \bibinfo{pages}{1863--1876}.

\bibitem[{Sun \emph{et~al.}(2015)Sun, Kang, Li, and Meng}]{sun2015voice}
\bibinfo{author}{L.~Sun}, \bibinfo{author}{S.~Kang}, \bibinfo{author}{K.~Li},
  \bibinfo{author}{H.~Meng}, \bibinfo{title}{Voice conversion using deep
  bidirectional long short-term memory based recurrent neural networks}, in:
  \bibinfo{booktitle}{2015 IEEE international conference on acoustics, speech
  and signal processing (ICASSP)}, \bibinfo{organization}{IEEE},
  \bibinfo{pages}{4869--4873}, \bibinfo{year}{2015}.

\bibitem[{Hsu \emph{et~al.}(2016)Hsu, Hwang, Wu, Tsao, and Wang}]{hsu2016voice}
\bibinfo{author}{C.-C. Hsu}, \bibinfo{author}{H.-T. Hwang},
  \bibinfo{author}{Y.-C. Wu}, \bibinfo{author}{Y.~Tsao}, \bibinfo{author}{H.-M.
  Wang}, \bibinfo{title}{Voice conversion from non-parallel corpora using
  variational auto-encoder}, in: \bibinfo{booktitle}{2016 Asia-Pacific Signal
  and Information Processing Association Annual Summit and Conference
  (APSIPA)}, \bibinfo{organization}{IEEE}, \bibinfo{pages}{1--6},
  \bibinfo{year}{2016}.

\bibitem[{Hsu \emph{et~al.}(2017{\natexlab{a}})Hsu, Hwang, Wu, Tsao, and
  Wang}]{hsu2017voice}
\bibinfo{author}{C.-C. Hsu}, \bibinfo{author}{H.-T. Hwang},
  \bibinfo{author}{Y.-C. Wu}, \bibinfo{author}{Y.~Tsao}, \bibinfo{author}{H.-M.
  Wang}, \bibinfo{title}{Voice conversion from unaligned corpora using
  variational autoencoding wasserstein generative adversarial networks},
  \bibinfo{journal}{arXiv preprint arXiv:1704.00849} .

\bibitem[{Kaneko and Kameoka(2018{\natexlab{a}})}]{kaneko2017parallel}
\bibinfo{author}{T.~Kaneko}, \bibinfo{author}{H.~Kameoka},
  \bibinfo{title}{Parallel-data-free voice conversion using cycle-consistent
  adversarial networks}, in: \bibinfo{booktitle}{26th European Signal
  Processing Conference (EUSIPCO)}, \bibinfo{pages}{2114--2118},
  \bibinfo{year}{2018}{\natexlab{a}}.

\bibitem[{Kaneko and Kameoka(2018{\natexlab{b}})}]{kaneko2018cyclegan}
\bibinfo{author}{T.~Kaneko}, \bibinfo{author}{H.~Kameoka},
  \bibinfo{title}{Cyclegan-vc: Non-parallel voice conversion using
  cycle-consistent adversarial networks}, in: \bibinfo{booktitle}{2018 26th
  European Signal Processing Conference (EUSIPCO)},
  \bibinfo{organization}{IEEE}, \bibinfo{pages}{2100--2104},
  \bibinfo{year}{2018}{\natexlab{b}}.

\bibitem[{Kameoka \emph{et~al.}(2018)Kameoka, Kaneko, Tanaka, and
  Hojo}]{kameoka2018stargan}
\bibinfo{author}{H.~Kameoka}, \bibinfo{author}{T.~Kaneko},
  \bibinfo{author}{K.~Tanaka}, \bibinfo{author}{N.~Hojo},
  \bibinfo{title}{Stargan-vc: Non-parallel many-to-many voice conversion using
  star generative adversarial networks}, in: \bibinfo{booktitle}{IEEE Spoken
  Language Technology Workshop (SLT)}, \bibinfo{organization}{IEEE},
  \bibinfo{pages}{266--273}, \bibinfo{year}{2018}.

\bibitem[{Tanaka \emph{et~al.}(2019)Tanaka, Kameoka, Kaneko, and
  Hojo}]{tanaka2019atts2s}
\bibinfo{author}{K.~Tanaka}, \bibinfo{author}{H.~Kameoka},
  \bibinfo{author}{T.~Kaneko}, \bibinfo{author}{N.~Hojo},
  \bibinfo{title}{AttS2S-VC: Sequence-to-sequence voice conversion with
  attention and context preservation mechanisms}, in: \bibinfo{booktitle}{IEEE
  International Conference on Acoustics, Speech and Signal Processing
  (ICASSP)}, \bibinfo{organization}{IEEE}, \bibinfo{pages}{6805--6809},
  \bibinfo{year}{2019}.

\bibitem[{Zhang \emph{et~al.}(2019)Zhang, Ling, Liu, Jiang, and
  Dai}]{zhang2019sequence}
\bibinfo{author}{J.-X. Zhang}, \bibinfo{author}{Z.-H. Ling},
  \bibinfo{author}{L.-J. Liu}, \bibinfo{author}{Y.~Jiang},
  \bibinfo{author}{L.-R. Dai}, \bibinfo{title}{Sequence-to-sequence acoustic
  modeling for voice conversion}, \bibinfo{journal}{IEEE/ACM Transactions on
  Audio, Speech, and Language Processing}
  \bibinfo{volume}{27}~(\bibinfo{number}{3}) (\bibinfo{year}{2019})
  \bibinfo{pages}{631--644}.

\bibitem[{Du \emph{et~al.}(2021)Du, Tian, Xie, and Li}]{du2021optimizing}
\bibinfo{author}{H.~Du}, \bibinfo{author}{X.~Tian}, \bibinfo{author}{L.~Xie},
  \bibinfo{author}{H.~Li}, \bibinfo{title}{Optimizing voice conversion network
  with cycle consistency loss of speaker identity}, in:
  \bibinfo{booktitle}{2021 IEEE Spoken Language Technology Workshop (SLT)},
  \bibinfo{organization}{IEEE}, \bibinfo{pages}{507--513},
  \bibinfo{year}{2021}.

\bibitem[{Wang \emph{et~al.}(2021{\natexlab{a}})Wang, Zhou, Yang, Li, Du, Xie,
  Gan, Chen, and Li}]{wang2021enriching}
\bibinfo{author}{Z.~Wang}, \bibinfo{author}{X.~Zhou},
  \bibinfo{author}{F.~Yang}, \bibinfo{author}{T.~Li}, \bibinfo{author}{H.~Du},
  \bibinfo{author}{L.~Xie}, \bibinfo{author}{W.~Gan},
  \bibinfo{author}{H.~Chen}, \bibinfo{author}{H.~Li}, \bibinfo{title}{Enriching
  Source Style Transfer in Recognition-Synthesis based Non-Parallel Voice
  Conversion}, \bibinfo{journal}{arXiv preprint arXiv:2106.08741} .

\bibitem[{Chou and Lee(2019)}]{chou2019one}
\bibinfo{author}{J.-c. Chou}, \bibinfo{author}{H.-Y. Lee},
  \bibinfo{title}{One-Shot Voice Conversion by Separating Speaker and Content
  Representations with Instance Normalization}, \bibinfo{journal}{Proc.
  Interspeech 2019}  (\bibinfo{year}{2019}) \bibinfo{pages}{664--668}.

\bibitem[{Qian \emph{et~al.}(2019)Qian, Zhang, Chang, Yang, and
  Hasegawa-Johnson}]{qian2019autovc}
\bibinfo{author}{K.~Qian}, \bibinfo{author}{Y.~Zhang},
  \bibinfo{author}{S.~Chang}, \bibinfo{author}{X.~Yang},
  \bibinfo{author}{M.~Hasegawa-Johnson}, \bibinfo{title}{Autovc: Zero-shot
  voice style transfer with only autoencoder loss}, in:
  \bibinfo{booktitle}{International Conference on Machine Learning},
  \bibinfo{organization}{PMLR}, \bibinfo{pages}{5210--5219},
  \bibinfo{year}{2019}.

\bibitem[{Du and Xie(2021)}]{du2021improving}
\bibinfo{author}{H.~Du}, \bibinfo{author}{L.~Xie}, \bibinfo{title}{Improving
  robustness of one-shot voice conversion with deep discriminative speaker
  encoder}, \bibinfo{journal}{arXiv preprint arXiv:2106.10406} .

\bibitem[{Wang \emph{et~al.}(2021{\natexlab{b}})Wang, Xie, Li, Du, Xie, Zhu,
  and Bi}]{wang2021one}
\bibinfo{author}{Z.~Wang}, \bibinfo{author}{Q.~Xie}, \bibinfo{author}{T.~Li},
  \bibinfo{author}{H.~Du}, \bibinfo{author}{L.~Xie}, \bibinfo{author}{P.~Zhu},
  \bibinfo{author}{M.~Bi}, \bibinfo{title}{One-shot Voice Conversion For Style
  Transfer Based On Speaker Adaptation}, \bibinfo{journal}{arXiv preprint
  arXiv:2111.12277} .

\bibitem[{Huang \emph{et~al.}(2021)Huang, Lin, and Lee}]{huang2021far}
\bibinfo{author}{T.-h. Huang}, \bibinfo{author}{J.-h. Lin},
  \bibinfo{author}{H.-y. Lee}, \bibinfo{title}{How Far Are We from Robust Voice
  Conversion: A Survey}, in: \bibinfo{booktitle}{2021 IEEE Spoken Language
  Technology Workshop (SLT)}, \bibinfo{organization}{IEEE},
  \bibinfo{pages}{514--521}, \bibinfo{year}{2021}.

\bibitem[{Chou \emph{et~al.}(2018)Chou, Yeh, Lee, and Lee}]{chou2018multi}
\bibinfo{author}{J.-c. Chou}, \bibinfo{author}{C.-c. Yeh},
  \bibinfo{author}{H.-y. Lee}, \bibinfo{author}{L.-s. Lee},
  \bibinfo{title}{Multi-target Voice Conversion without Parallel Data by
  Adversarially Learning Disentangled Audio Representations},
  \bibinfo{journal}{Proc. Interspeech 2018}  (\bibinfo{year}{2018})
  \bibinfo{pages}{501--505}.

\bibitem[{Li \emph{et~al.}(2018)Li, Wang, Chen, Shi, Tang, and
  Zheng}]{li2018deep}
\bibinfo{author}{L.~Li}, \bibinfo{author}{D.~Wang}, \bibinfo{author}{Y.~Chen},
  \bibinfo{author}{Y.~Shi}, \bibinfo{author}{Z.~Tang}, \bibinfo{author}{T.~F.
  Zheng}, \bibinfo{title}{Deep factorization for speech signal}, in:
  \bibinfo{booktitle}{2018 IEEE International Conference on Acoustics, Speech
  and Signal Processing (ICASSP)}, \bibinfo{organization}{IEEE},
  \bibinfo{pages}{5094--5098}, \bibinfo{year}{2018}.

\bibitem[{Sun \emph{et~al.}(2017)Sun, Zhang, Xie, and
  Zhang}]{sun2017unsupervised}
\bibinfo{author}{S.~Sun}, \bibinfo{author}{B.~Zhang}, \bibinfo{author}{L.~Xie},
  \bibinfo{author}{Y.~Zhang}, \bibinfo{title}{An unsupervised deep domain
  adaptation approach for robust speech recognition},
  \bibinfo{journal}{Neurocomputing} \bibinfo{volume}{257}
  (\bibinfo{year}{2017}) \bibinfo{pages}{79--87}.

\bibitem[{Wang \emph{et~al.}(2018)Wang, Rao, Sun, Xie, Chng, and
  Li}]{wang2018unsupervised}
\bibinfo{author}{Q.~Wang}, \bibinfo{author}{W.~Rao}, \bibinfo{author}{S.~Sun},
  \bibinfo{author}{L.~Xie}, \bibinfo{author}{E.~S. Chng},
  \bibinfo{author}{H.~Li}, \bibinfo{title}{Unsupervised domain adaptation via
  domain adversarial training for speaker recognition}, in:
  \bibinfo{booktitle}{2018 IEEE International Conference on Acoustics, Speech
  and Signal Processing (ICASSP)}, \bibinfo{organization}{IEEE},
  \bibinfo{pages}{4889--4893}, \bibinfo{year}{2018}.

\bibitem[{Takashima \emph{et~al.}(2014)Takashima, Aihara, Takiguchi, and
  Ariki}]{aihara2014noise}
\bibinfo{author}{R.~Takashima}, \bibinfo{author}{R.~Aihara},
  \bibinfo{author}{T.~Takiguchi}, \bibinfo{author}{Y.~Ariki},
  \bibinfo{title}{Noise-robust voice conversion based on sparse spectral
  mapping using non-negative matrix factorization}, \bibinfo{journal}{IEICE
  TRANSACTIONS on Information and Systems}
  \bibinfo{volume}{97}~(\bibinfo{number}{6}) (\bibinfo{year}{2014})
  \bibinfo{pages}{1411--1418}.

\bibitem[{Aihara \emph{et~al.}(2015)Aihara, Fujii, Nakashika, Takiguchi, and
  Ariki}]{aihara2015small}
\bibinfo{author}{R.~Aihara}, \bibinfo{author}{T.~Fujii},
  \bibinfo{author}{T.~Nakashika}, \bibinfo{author}{T.~Takiguchi},
  \bibinfo{author}{Y.~Ariki}, \bibinfo{title}{Small-parallel exemplar-based
  voice conversion in noisy environments using affine non-negative matrix
  factorization}, \bibinfo{journal}{EURASIP Journal on Audio, Speech, and Music
  Processing} \bibinfo{volume}{2015}~(\bibinfo{number}{1})
  (\bibinfo{year}{2015}) \bibinfo{pages}{1--9}.

\bibitem[{Hsu \emph{et~al.}(2017{\natexlab{b}})Hsu, Zhang, and
  Glass}]{hsu2017unsupervised}
\bibinfo{author}{W.-N. Hsu}, \bibinfo{author}{Y.~Zhang},
  \bibinfo{author}{J.~Glass}, \bibinfo{title}{Unsupervised learning of
  disentangled and interpretable representations from sequential data}, in:
  \bibinfo{booktitle}{Advances in neural information processing systems},
  \bibinfo{pages}{1878--1889}, \bibinfo{year}{2017}{\natexlab{b}}.

\bibitem[{Valentini-Botinhao \emph{et~al.}(2016)Valentini-Botinhao, Wang,
  Takaki, and Yamagishi}]{valentini2016investigating}
\bibinfo{author}{C.~Valentini-Botinhao}, \bibinfo{author}{X.~Wang},
  \bibinfo{author}{S.~Takaki}, \bibinfo{author}{J.~Yamagishi},
  \bibinfo{title}{Investigating RNN-based speech enhancement methods for
  noise-robust Text-to-Speech.}, in: \bibinfo{booktitle}{SSW},
  \bibinfo{pages}{146--152}, \bibinfo{year}{2016}.

\bibitem[{Yang \emph{et~al.}(2020)Yang, Wang, and Xie}]{yang2020adversarial}
\bibinfo{author}{S.~Yang}, \bibinfo{author}{Y.~Wang}, \bibinfo{author}{L.~Xie},
  \bibinfo{title}{Adversarial feature learning and unsupervised clustering
  based speech synthesis for found data with acoustic and textual noise},
  \bibinfo{journal}{IEEE Signal Processing Letters} \bibinfo{volume}{27}
  (\bibinfo{year}{2020}) \bibinfo{pages}{1730--1734}.

\bibitem[{Sekkate \emph{et~al.}(2019)Sekkate, Khalil, Adib, and
  Ben~Jebara}]{sekkate2019investigation}
\bibinfo{author}{S.~Sekkate}, \bibinfo{author}{M.~Khalil},
  \bibinfo{author}{A.~Adib}, \bibinfo{author}{S.~Ben~Jebara},
  \bibinfo{title}{An Investigation of a Feature-Level Fusion for Noisy Speech
  Emotion Recognition}, \bibinfo{journal}{Computers}
  \bibinfo{volume}{8}~(\bibinfo{number}{4}) (\bibinfo{year}{2019})
  \bibinfo{pages}{91}.

\bibitem[{Hsu \emph{et~al.}(2019)Hsu, Zhang, Weiss, Chung, Wang, Wu, and
  Glass}]{hsu2019disentangling}
\bibinfo{author}{W.-N. Hsu}, \bibinfo{author}{Y.~Zhang}, \bibinfo{author}{R.~J.
  Weiss}, \bibinfo{author}{Y.-A. Chung}, \bibinfo{author}{Y.~Wang},
  \bibinfo{author}{Y.~Wu}, \bibinfo{author}{J.~Glass},
  \bibinfo{title}{Disentangling correlated speaker and noise for speech
  synthesis via data augmentation and adversarial factorization}, in:
  \bibinfo{booktitle}{IEEE International Conference on Acoustics, Speech and
  Signal Processing (ICASSP)}, \bibinfo{organization}{IEEE},
  \bibinfo{pages}{5901--5905}, \bibinfo{year}{2019}.

\bibitem[{Ganin \emph{et~al.}(2016)Ganin, Ustinova, Ajakan, Germain,
  Larochelle, Laviolette, Marchand, and Lempitsky}]{ganin2016domain}
\bibinfo{author}{Y.~Ganin}, \bibinfo{author}{E.~Ustinova},
  \bibinfo{author}{H.~Ajakan}, \bibinfo{author}{P.~Germain},
  \bibinfo{author}{H.~Larochelle}, \bibinfo{author}{F.~Laviolette},
  \bibinfo{author}{M.~Marchand}, \bibinfo{author}{V.~Lempitsky},
  \bibinfo{title}{Domain-adversarial training of neural networks},
  \bibinfo{journal}{The Journal of Machine Learning Research}
  \bibinfo{volume}{17}~(\bibinfo{number}{1}) (\bibinfo{year}{2016})
  \bibinfo{pages}{2096--2030}.

\bibitem[{Shinohara(2016)}]{shinohara2016adversarial}
\bibinfo{author}{Y.~Shinohara}, \bibinfo{title}{Adversarial Multi-Task Learning
  of Deep Neural Networks for Robust Speech Recognition.}, in:
  \bibinfo{booktitle}{Interspeech}, \bibinfo{organization}{San Francisco, CA,
  USA}, \bibinfo{pages}{2369--2372}, \bibinfo{year}{2016}.

\bibitem[{Tu \emph{et~al.}(2019)Tu, Mak, and Chien}]{tu2019variational}
\bibinfo{author}{Y.~Tu}, \bibinfo{author}{M.-W. Mak}, \bibinfo{author}{J.-T.
  Chien}, \bibinfo{title}{Variational Domain Adversarial Learning for Speaker
  Verification.}, in: \bibinfo{booktitle}{Interspeech},
  \bibinfo{pages}{4315--4319}, \bibinfo{year}{2019}.

\bibitem[{Liao \emph{et~al.}(2018)Liao, Tsao, Lee, and Wang}]{liao2018noise}
\bibinfo{author}{C.-F. Liao}, \bibinfo{author}{Y.~Tsao}, \bibinfo{author}{H.-Y.
  Lee}, \bibinfo{author}{H.-M. Wang}, \bibinfo{title}{Noise adaptive speech
  enhancement using domain adversarial training}, \bibinfo{journal}{arXiv
  preprint arXiv:1807.07501} .

\bibitem[{Lim \emph{et~al.}(2020)Lim, Kim, and Kim}]{lim2020cross}
\bibinfo{author}{H.~Lim}, \bibinfo{author}{Y.~Kim}, \bibinfo{author}{H.~Kim},
  \bibinfo{title}{Cross-Informed Domain Adversarial Training for Noise-Robust
  Wake-Up Word Detection}, \bibinfo{journal}{IEEE Signal Processing Letters}
  \bibinfo{volume}{27} (\bibinfo{year}{2020}) \bibinfo{pages}{1769--1773}.

\bibitem[{Okabe \emph{et~al.}(2018)Okabe, Koshinaka, and
  Shinoda}]{okabe2018attentive}
\bibinfo{author}{K.~Okabe}, \bibinfo{author}{T.~Koshinaka},
  \bibinfo{author}{K.~Shinoda}, \bibinfo{title}{Attentive Statistics Pooling
  for Deep Speaker Embedding}, \bibinfo{journal}{Proc. Interspeech 2018}
  (\bibinfo{year}{2018}) \bibinfo{pages}{2252--2256}.

\bibitem[{Tang \emph{et~al.}(2019)Tang, Ding, Huang, He, and
  Zhou}]{tang2019deep}
\bibinfo{author}{Y.~Tang}, \bibinfo{author}{G.~Ding},
  \bibinfo{author}{J.~Huang}, \bibinfo{author}{X.~He},
  \bibinfo{author}{B.~Zhou}, \bibinfo{title}{Deep speaker embedding learning
  with multi-level pooling for text-independent speaker verification}, in:
  \bibinfo{booktitle}{ICASSP 2019-2019 IEEE International Conference on
  Acoustics, Speech and Signal Processing (ICASSP)},
  \bibinfo{organization}{IEEE}, \bibinfo{pages}{6116--6120},
  \bibinfo{year}{2019}.

\bibitem[{Mor \emph{et~al.}(2018)Mor, Wolf, Polyak, and
  Taigman}]{mor2018universal}
\bibinfo{author}{N.~Mor}, \bibinfo{author}{L.~Wolf},
  \bibinfo{author}{A.~Polyak}, \bibinfo{author}{Y.~Taigman}, \bibinfo{title}{A
  universal music translation network}, \bibinfo{journal}{arXiv preprint
  arXiv:1805.07848} .

\bibitem[{Ulyanov \emph{et~al.}(2016)Ulyanov, Vedaldi, and
  Lempitsky}]{ulyanov2016instance}
\bibinfo{author}{D.~Ulyanov}, \bibinfo{author}{A.~Vedaldi},
  \bibinfo{author}{V.~Lempitsky}, \bibinfo{title}{Instance normalization: The
  missing ingredient for fast stylization}, \bibinfo{journal}{arXiv preprint
  arXiv:1607.08022} .

\bibitem[{Locatello \emph{et~al.}(2019)Locatello, Bauer, Lucic, Raetsch, Gelly,
  Sch{\"o}lkopf, and Bachem}]{locatello2019challenging}
\bibinfo{author}{F.~Locatello}, \bibinfo{author}{S.~Bauer},
  \bibinfo{author}{M.~Lucic}, \bibinfo{author}{G.~Raetsch},
  \bibinfo{author}{S.~Gelly}, \bibinfo{author}{B.~Sch{\"o}lkopf},
  \bibinfo{author}{O.~Bachem}, \bibinfo{title}{Challenging common assumptions
  in the unsupervised learning of disentangled representations}, in:
  \bibinfo{booktitle}{international conference on machine learning},
  \bibinfo{organization}{PMLR}, \bibinfo{pages}{4114--4124},
  \bibinfo{year}{2019}.

\bibitem[{Lu \emph{et~al.}(2013)Lu, Tsao, Matsuda, and Hori}]{lu2013speech}
\bibinfo{author}{X.~Lu}, \bibinfo{author}{Y.~Tsao},
  \bibinfo{author}{S.~Matsuda}, \bibinfo{author}{C.~Hori},
  \bibinfo{title}{Speech enhancement based on deep denoising autoencoder.}, in:
  \bibinfo{booktitle}{Interspeech}, vol. \bibinfo{volume}{2013},
  \bibinfo{pages}{436--440}, \bibinfo{year}{2013}.

\bibitem[{Shivakumar and Georgiou(2016)}]{shivakumar2016perception}
\bibinfo{author}{P.~G. Shivakumar}, \bibinfo{author}{P.~G. Georgiou},
  \bibinfo{title}{Perception optimized deep denoising autoencoders for speech
  enhancement.}, in: \bibinfo{booktitle}{Interspeech},
  \bibinfo{pages}{3743--3747}, \bibinfo{year}{2016}.

\bibitem[{Veaux \emph{et~al.}(2017)Veaux, Yamagishi, MacDonald
  \emph{et~al.}}]{veaux2017cstr}
\bibinfo{author}{C.~Veaux}, \bibinfo{author}{J.~Yamagishi},
  \bibinfo{author}{K.~MacDonald}, \emph{et~al.}, \bibinfo{title}{CSTR VCTK
  corpus: English multi-speaker corpus for CSTR voice cloning toolkit},
  \bibinfo{journal}{University of Edinburgh. The Centre for Speech Technology
  Research (CSTR)} .

\bibitem[{Vincent \emph{et~al.}(2017)Vincent, Watanabe, Nugraha, Barker, and
  Marxer}]{vincent2017analysis}
\bibinfo{author}{E.~Vincent}, \bibinfo{author}{S.~Watanabe},
  \bibinfo{author}{A.~A. Nugraha}, \bibinfo{author}{J.~Barker},
  \bibinfo{author}{R.~Marxer}, \bibinfo{title}{An analysis of environment,
  microphone and data simulation mismatches in robust speech recognition},
  \bibinfo{journal}{Computer Speech \& Language} \bibinfo{volume}{46}
  (\bibinfo{year}{2017}) \bibinfo{pages}{535--557}.

\bibitem[{Kominek and Black(2004)}]{kominek2004cmu}
\bibinfo{author}{J.~Kominek}, \bibinfo{author}{A.~W. Black},
  \bibinfo{title}{The CMU Arctic speech databases}, in:
  \bibinfo{booktitle}{Fifth ISCA workshop on speech synthesis},
  \bibinfo{year}{2004}.

\bibitem[{Varga and Steeneken(1993)}]{varga1993assNOISEessment}
\bibinfo{author}{A.~Varga}, \bibinfo{author}{H.~J. Steeneken},
  \bibinfo{title}{Assessment for automatic speech recognition: II. NOISEX-92: A
  database and an experiment to study the effect of additive noise on speech
  recognition systems}, \bibinfo{journal}{Speech communication}
  \bibinfo{volume}{12}~(\bibinfo{number}{3}) (\bibinfo{year}{1993})
  \bibinfo{pages}{247--251}.

\bibitem[{Botinhao \emph{et~al.}(2016)Botinhao, Wang, Takaki, and
  Yamagishi}]{botinhao2016investigating}
\bibinfo{author}{C.~V. Botinhao}, \bibinfo{author}{X.~Wang},
  \bibinfo{author}{S.~Takaki}, \bibinfo{author}{J.~Yamagishi},
  \bibinfo{title}{Investigating RNN-based speech enhancement methods for
  noise-robust Text-to-Speech}, in: \bibinfo{booktitle}{9th ISCA Speech
  Synthesis Workshop}, \bibinfo{pages}{159--165}, \bibinfo{year}{2016}.

\bibitem[{McFee \emph{et~al.}(2015)McFee, Raffel, Liang, Ellis, McVicar,
  Battenberg, and Nieto}]{mcfee2015librosa}
\bibinfo{author}{B.~McFee}, \bibinfo{author}{C.~Raffel},
  \bibinfo{author}{D.~Liang}, \bibinfo{author}{D.~P. Ellis},
  \bibinfo{author}{M.~McVicar}, \bibinfo{author}{E.~Battenberg},
  \bibinfo{author}{O.~Nieto}, \bibinfo{title}{librosa: Audio and music signal
  analysis in python}, in: \bibinfo{booktitle}{Proceedings of the 14th python
  in science conference}, vol.~\bibinfo{volume}{8}, \bibinfo{pages}{18--25},
  \bibinfo{year}{2015}.

\bibitem[{Yamamoto \emph{et~al.}(2020)Yamamoto, Song, and
  Kim}]{yamamoto2020parallel}
\bibinfo{author}{R.~Yamamoto}, \bibinfo{author}{E.~Song},
  \bibinfo{author}{J.-M. Kim}, \bibinfo{title}{Parallel {WaveGAN}: A fast
  waveform generation model based on generative adversarial networks with
  multi-resolution spectrogram}, in: \bibinfo{booktitle}{International
  Conference on Acoustics, Speech and Signal Processing},
  \bibinfo{organization}{IEEE}, \bibinfo{pages}{6199--6203},
  \bibinfo{year}{2020}.

\bibitem[{Hu \emph{et~al.}(2020)Hu, Liu, Lv, Xing, Zhang, Fu, Wu, Zhang, and
  Xie}]{hudccrn}
\bibinfo{author}{Y.~Hu}, \bibinfo{author}{Y.~Liu}, \bibinfo{author}{S.~Lv},
  \bibinfo{author}{M.~Xing}, \bibinfo{author}{S.~Zhang},
  \bibinfo{author}{Y.~Fu}, \bibinfo{author}{J.~Wu}, \bibinfo{author}{B.~Zhang},
  \bibinfo{author}{L.~Xie}, \bibinfo{title}{DCCRN: Deep Complex Convolution
  Recurrent Network for Phase-Aware Speech Enhancement.}, in:
  \bibinfo{booktitle}{Interspeech}, \bibinfo{year}{2020}.

\bibitem[{Kingma and Ba(2014)}]{2014Adam}
\bibinfo{author}{D.~Kingma}, \bibinfo{author}{J.~Ba}, \bibinfo{title}{Adam: A
  Method for Stochastic Optimization}, \bibinfo{journal}{Computer Science} .

\bibitem[{Machado and Queiroz(2010)}]{machado2010voice}
\bibinfo{author}{A.~F. Machado}, \bibinfo{author}{M.~Queiroz},
  \bibinfo{title}{Voice conversion: A critical survey}, \bibinfo{journal}{Proc.
  Sound and Music Computing (SMC)}  (\bibinfo{year}{2010})
  \bibinfo{pages}{1--8}.

\bibitem[{Rix \emph{et~al.}(2001)Rix, Beerends, Hollier, and
  Hekstra}]{rix2001perceptual}
\bibinfo{author}{A.~W. Rix}, \bibinfo{author}{J.~G. Beerends},
  \bibinfo{author}{M.~P. Hollier}, \bibinfo{author}{A.~P. Hekstra},
  \bibinfo{title}{Perceptual evaluation of speech quality (PESQ)-a new method
  for speech quality assessment of telephone networks and codecs}, in:
  \bibinfo{booktitle}{2001 IEEE International Conference on Acoustics, Speech,
  and Signal Processing. Proceedings (Cat. No. 01CH37221)},
  vol.~\bibinfo{volume}{2}, \bibinfo{organization}{IEEE},
  \bibinfo{pages}{749--752}, \bibinfo{year}{2001}.

\bibitem[{Gulati \emph{et~al.}(2020)Gulati, Qin, Chiu, Parmar, Zhang, Yu, Han,
  Wang, Zhang, Wu \emph{et~al.}}]{gulati2020conformer}
\bibinfo{author}{A.~Gulati}, \bibinfo{author}{J.~Qin}, \bibinfo{author}{C.-C.
  Chiu}, \bibinfo{author}{N.~Parmar}, \bibinfo{author}{Y.~Zhang},
  \bibinfo{author}{J.~Yu}, \bibinfo{author}{W.~Han}, \bibinfo{author}{S.~Wang},
  \bibinfo{author}{Z.~Zhang}, \bibinfo{author}{Y.~Wu}, \emph{et~al.},
  \bibinfo{title}{Conformer: Convolution-augmented transformer for speech
  recognition}, \bibinfo{journal}{arXiv preprint arXiv:2005.08100} .

\bibitem[{Laurens and Hinton(2008)}]{2008Visualizing}
\bibinfo{author}{V.~D.~M. Laurens}, \bibinfo{author}{G.~Hinton},
  \bibinfo{title}{Visualizing Data using t-SNE}, \bibinfo{journal}{Journal of
  Machine Learning Research} \bibinfo{volume}{9}~(\bibinfo{number}{2605})
  (\bibinfo{year}{2008}) \bibinfo{pages}{2579--2605}.

\end{thebibliography}

\end{document}